\documentclass[prl,twocolumn,showpacs,amsmath,amssymb,superscriptaddress]{revtex4-1}

\usepackage{color}
\usepackage{amssymb}
\usepackage{amsthm}
\usepackage{graphicx}
\usepackage{epsfig}
\usepackage{subfigure}
\usepackage{amsmath,amssymb}
\usepackage[colorlinks=true,citecolor=blue,urlcolor=blue]{hyperref}
\usepackage{color}
\usepackage{bm}
\newcommand{\be}{\begin{equation}}
\newcommand{\ee}{\end{equation}}
\newcommand{\ba}{\begin{eqnarray}}
\newcommand{\ea}{\end{eqnarray}}
\newcommand{\ban}{\begin{eqnarray*}}
\newcommand{\ean}{\end{eqnarray*}}
\newcommand{\ket}[1]{\mbox{$ | #1 \rangle $}}
\newcommand{\bra}[1]{\mbox{$ \langle #1 | $}}

\newcommand{\one}{\leavevmode\hbox{\small1\normalsize\kern-.33em1}}

\begin{document}

\title{\Large\textbf{Detecting the entanglement of vortices in ultracold bosons\\ with artificial gauge fields}}
\author{Li Dai}
\affiliation{Beijing National Laboratory for Condensed Matter Physics, Institute of Physics, Chinese Academy of Sciences, Beijing 100190, China}

\author{Lin Xia}
\affiliation{Beijing National Laboratory for Condensed Matter Physics, Institute of Physics, Chinese Academy of Sciences, Beijing 100190, China}

\author{Lin Zhuang}
\affiliation{School of Physics, Sun Yat-Sen University, Guangzhou 510275, China}

\author{Wu-Ming Liu}
\email{wmliu@iphy.ac.cn}
\affiliation{Beijing National Laboratory for Condensed Matter Physics, Institute of Physics, Chinese Academy of Sciences, Beijing 100190, China}

\begin{abstract}
The entanglement of vortices in a two-dimensional Bose-Hubbard model with artificial gauge fields is investigated using the exact diagonalization techniques. We propose an effective Hamiltonian for the spin-spin interactions between vortices responsible for this entanglement, and show that the entanglement can be detected through the quantum interference of the bosons in the vortex centers achieved using the Raman coupling and the quantum gas microscope. The strong bosonic coherence between the vortex centers originates from the charge-density wave order in the vortex core. It is robust against the varying of the pinning strength for the vortices to a wide range, and the coherent bosons can be viewed as a qubit stored in the ground state of the system. Our proposal provides a feasible scheme of quantum memory for storing qubits useful in quantum computation. 
\end{abstract} 

\date{\today}
\maketitle

\emph{Introduction}. The vortex describes the circulation of some physical quantities of a system. For a strongly correlated many-body system, the vortex may have a rich internal structure due to the coexistence of multiple orders~\cite{PhysRevA.69.043609,PhysRevA.79.021602,PhysRevB.97.174510,PhysRevB.97.174511}. For instance, for the hard-core bosons in the two-dimensional lattice at half filling, the vortex has no density depletion in the core where the charge-density wave (CDW) order develops, competing with the superfluid order~\cite{PhysRevB.82.134510}. This is in contrast to the conventional vortex where the core has a large density depletion, e.g. in a Bose-Einstein condensate (BEC) described by the Gross-Pitaevskii (GP) theory~\cite{Ginzburg_Pitaevskii,Gross1961,Pitaevskii1961}. Due to the charge-vortex duality~\cite{PhysRevB.39.2756,PhysRevLett.63.903,duality2,PhysRevLett.113.240601}, the CDW acts as an effective magnetic field for the vortex to circulate in the dual lattice. As a result, the vortex carries an internal spin-half degree of freedom~\cite{PhysRevLett.102.070403} which denotes the counterclockwise (spin up $\uparrow$) and clockwise (spin down $\downarrow$) vortex currents~\cite{pnas}. When the system is penetrated with two flux quanta, there are two vortices located far from each other, yet with their spins entangled~\cite{arxiv0701571v1}.

The vortex spin and the entangled spin state are theoretical concepts introduced to facilitate the understanding of the quantum dynamics of vortices~\cite{PhysRevLett.102.070403,PhysRevB.82.134510,arxiv0701571v1}. It remains a question whether or not the vortex spin is a quantity with physical reality. In particular, how do the vortex spins interact with each other to create entanglement, and can this entanglement be detected in experiments? In this Letter, we propose an effective Hamiltonian for the spin-spin interactions between vortices responsible for this entanglement, and show that the entanglement can be detected through the interference of the bosons in the two vortex centers achieved using the Raman coupling~\cite{Raman1928} and the quantum gas microscope~\cite{Greiner_Nature,Bakr547}. The strong bosonic coherence between the vortex centers originates from the CDW order in the vortex core where the superfluid order is suppressed~\cite{PhysRevA.69.043609}. % The CDW order is absent in the vortex of the BEC described by the GP theory. 

The vortex entanglement has been studied in different contexts, including the high-temperature superconductors~\cite{PhysRevLett.71.3545,PhysRevLett.101.027003,PhysRevLett.60.1973,PhysRevLett.75.1380}, neutron stars~\cite{Neutron-Stars}, the fast-rotating BEC~\cite{PhysRevLett.87.120405,PhysRevA.83.013620}, two optically coupled BECs~\cite{PhysRevA.81.053625}, etc. For the former two cases, the vortex entanglement (or the braided flux) is induced either by the splayed defects or by thermal fluctuations. %: the thermally excited vortex lines wander and collide with each other forming an entangled flux liquid. 
For the fast-rotating BEC, the strongly correlated vortex liquid is formed due to the quantum fluctuations of the high-density vortices whose kinetic energy dominates over the Coulomb interactions. The last case is concerning the Fock-state entanglement of vortices with opposite angular momenta. The entanglement studied in this Letter is different from the above cases in that the degree of freedom for the entanglement is the vortex spin which emerges due to the quantum fluctuations of the CDW in the vortex core.

Detecting the entanglement of vortices in experiments will not only demonstrate the existence of the vortex spin and the CDW order in the vortex core, but also show the possibility of macroscopic entanglement in the strongly correlated many-body system. The macroscopic entanglement is a fundamental concept for testing the Bohr correspondence principle~\cite{Vedral_macro,Quantumness1,Quantumness2} and has important applications in quantum information processing~\cite{PhysRevA.59.1829,cats2,cats3}. The typical state with a macroscopic entanglement is in the form of two entangled Schr\"odinger cats~\cite{Schrödinger1935}: $\ket{\Psi_c}=\frac{1}{\sqrt 2}(\ket{D_1}\ket{L_2}+\ket{L_1}\ket{D_2})$, where $\ket{D_j}$ ($\ket{L_j}$) denotes that the cat $j$ is dead (alive). We notice that such a macroscopic entanglement is usually present in the ground state of the interacting systems in a double-well potential~\cite{PhysRevA.57.1208,Ho2004,PhysRevLett.98.060403}, or can be generated dynamically through atom-photon interactions~\cite{PhysRevLett.83.1319,etg_Nature,Julsgaard_Nature,PhysRevA.67.013607,PhysRevLett.90.030402,Oberthaler_Nature}. However, it is not expected to exist in a generic strongly correlated many-body system. This is because the ground state of such a system is a highly complicated many-body state for which a macroscopic entangled state in the simple form of the two entangled Schr\"odinger cats (a two-body state) is not likely to be present.

Nevertheless, due to the charge-vortex duality~\cite{PhysRevB.39.2756,PhysRevLett.63.903,duality2,PhysRevLett.113.240601} as mentioned earlier, the complicated strongly correlated many-body system with artificial gauge fields can be mapped to an effective system of interacting vortices. The latter can be a quantum few-body system if the gauge field is weak (as the number of vortices is small). This enables the observation of the macroscopic entanglement of vortices.

\begin{figure}[b]
%\begin{center}
%\vspace{10cm}
\subfigure[]{\label{Fig-1-optical_lattice}
\includegraphics[width=1.8in]{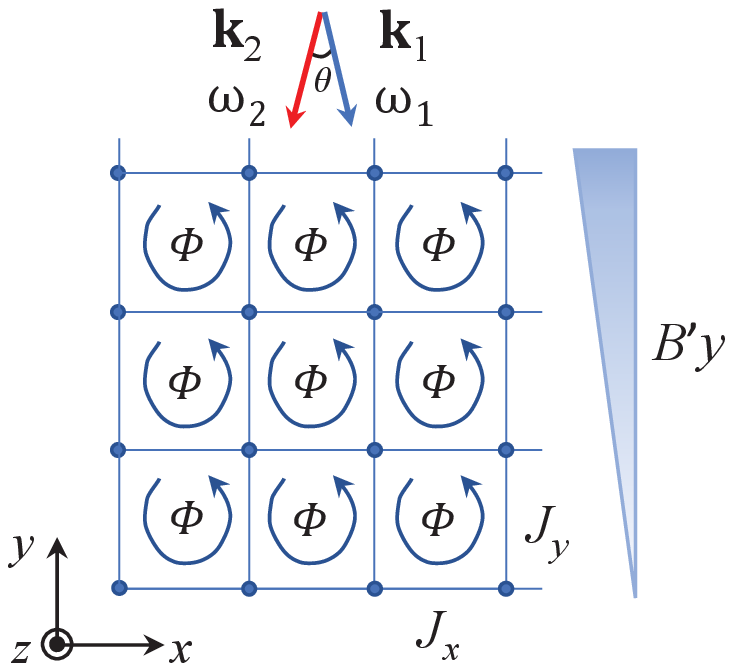}}
%\vspace{3mm}
\subfigure[]{\label{Fig-symmetry-axis}
\includegraphics[width=1in]{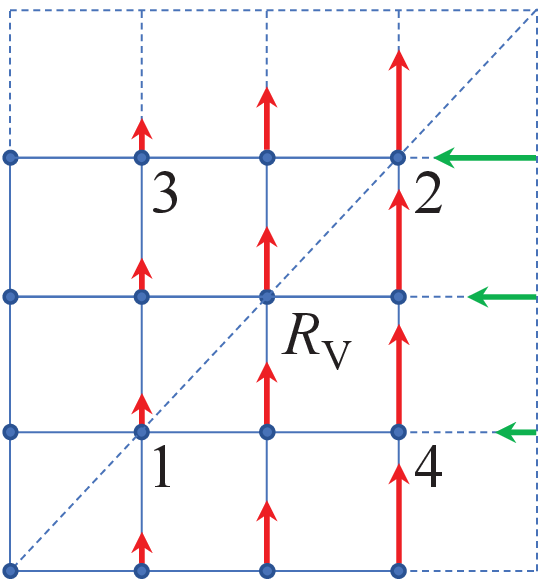}}
%\end{center}
\caption{(Color online). (a) The schematic of the two-dimensional Bose-Hubbard model on the square optical lattice. The direct tunneling along $x$ is $J_x$, while the tunneling along $y$ is inhibited by a magnetic field gradient $B'$. The two lasers with the frequencies $\omega_1,\omega_2$ and the wave vectors $\bm k_1,\bm k_2$ are applied at the small relative angle $\theta$ along the $-y$ axis, which induce a Raman coupling with the strength $J_y$ between the neighboring sites along $y$. This creates a uniform flux $\Phi$ in each plaquette. (b) The $4\times4$ square lattice with periodic boundary conditions. The anti-diagonal dashed line connecting the two vortex centers $1$ and $2$ denotes the axis of the reflection operation. The point $R_V$ is the vorticity center~\cite{PhysRevB.82.134510}. The green and red arrows denote the $x$ and $y$ components of the vector potential $\bm{A}$ corresponding to the flux $\Phi=\oint_\square\bm A\cdot d\bm l$ (``$\square$" denotes the plaquette).}
\end{figure}

\begin{figure}[b]
%\begin{center}
%\vspace{10cm}
%\subfigure[]{\label{Fig-symmetry-axis}
%\includegraphics[width=1.1in]{Fig-symmetry_axis}}
%\vspace{3mm}
%\subfigure[]{
\includegraphics[width=2.8in]{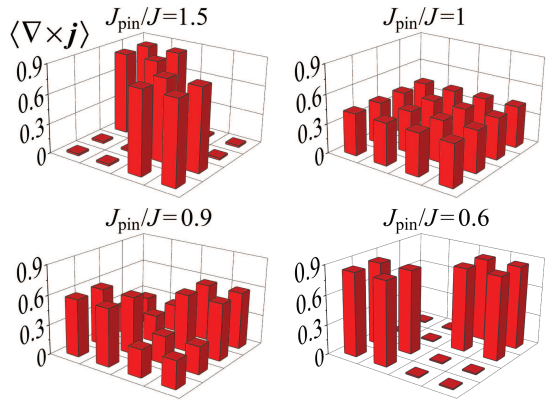}
\caption{(Color online). The vorticity of the electric current density $\langle\nabla\times\bm j\rangle$ of each plaquette in the $4\times4$ square lattice is plotted for different pinning strength $J_{\textrm{pin}}/J=1.5,1,0.9,0.6$ where the background vorticity is subtracted~\cite{supp,vorticity_deg}.}\label{Fig-vtc}
\end{figure}

\emph{Model}. We consider the two-dimensional Bose-Hubbard model on the square lattice with the Hamiltonian 
\ba\label{Hamiltonian}
H=&-&J\sum_{\langle\bm{rr'}\rangle}(e^{iA_{\bm{rr'}}}a^{\dagger}_{\bm{r}}a_{\bm{r'}}+\textrm{H.c.})+U\sum_{\bm{r}}n_{\bm{r}}(n_{\bm{r}}-1)\nonumber\\
&+&V\sum_{\langle\bm{rr'}\rangle}n_{\bm{r}}n_{\bm{r'}}-\mu\sum_{\bm{r}}n_{\bm{r}},
\ea
where $a_{\bm{r}}$($a^{\dagger}_{\bm{r}}$) is the annihilation (creation) operator for the bosons at the lattice site $\bm{r}$, $``\textrm{H.c.}"$ denotes the Hermitian conjugate of its previous term, $J$ is the hopping strength for the neighboring sites, $U$ ($V$) describes the on-site (nearest-neighbor) interactions, $\mu$ is the chemical potential, and $A_{\bm{rr'}}=\frac{q}{\hbar}\int_{\bm{r}}^{\bm{r'}}\bm{A}\cdot d\bm{l}$ is the lattice gauge field. Here, $q$ is the boson charge which is assumed to be $-e$ without loss of generality. For neutral bosons with artificial gauge fields, $q$ is only a parameter introduced to facilitate the physical description. In this way, $A_{\bm{rr'}}=\frac{2\pi}{\Phi_0}\int_{\bm{r'}}^{\bm{r}}\bm{A}\cdot d\bm{l}$ ($\Phi_0=h/e$ is the flux quantum). The gauge field corresponds to a magnetic field $\bm{B}=\nabla\times\bm{A}$ which creates a vortex configuration with the density $n_v=B/\Phi_0$~\cite{PhysRevLett.102.070403,PhysRevB.82.134510}.

The Hamiltonian can be realized in experiments with ultracold bosons in optical lattices~\cite{Jaksch_Zoller,Gerbier_Dalibard,PhysRevLett.94.086803,PhysRevLett.111.185301,PhysRevLett.111.185302,Lin_Nature,Aidelsburger}, the Josephson junction arrays~\cite{PhysRevLett.81.4484,FAZIO2001235,MACHIDA201344} and the high-temperature superconductors~\cite{PhysRevLett.63.903,PhysRevB.71.144508,Hoffman466}. The scheme of the optical lattice is shown in Fig.~\ref{Fig-1-optical_lattice}. The $^{87}$Rb atoms in the angular momentum eigenstate $\ket{F=1,m_F=-1}$ are loaded into a three-dimensional optical lattice, where the lattice depth along $z$ is sufficiently deep to inhibit the tunneling along $z$. This essentially creates a two-dimensional optical lattice. A magnetic field gradient $B'$ is applied along $y$ to generate an energy offset $\Delta=\mu_BB'a/2$ between the neighboring atoms along $y$, where $\mu_B$ is the Bohr magneton and $a$ is the lattice constant. This inhibits the direct tunneling along $y$, which is instead induced by applying two Raman lasers with the frequencies $\omega_1,\omega_2$ and the wave vectors $\bm k_1,\bm k_2$ at a small relative angle $\theta$ along the $-y$ axis ($\omega_1-\omega_2=\Delta/\hbar$). The corresponding tunneling strength $J_y=Je^{iA_{\bm{rr'}}}$ with $A_{\bm{rr'}}=(\bm k_1-\bm k_2)\cdot\bm{r'}$ for the bosons hopping from $\bm{r'}$ to $\bm r=\bm{r'}+\bm e_y$ with $\bm e_y$ being the unit vector along $y$~\cite{PhysRevLett.111.185301,PhysRevLett.111.185302}. The Raman lasers creates a uniform flux $\Phi=\delta k_x a\Phi_0/(2\pi)$ in each plaquette ($\delta k_x\approx2k\sin\theta$, $|\bm k_1|\approx|\bm k_2|=k$). The flux $\Phi$ can be tuned by adjusting the orientations of $\bm k_1,\bm k_2$.

The vortex spin denotes the two forms in which the vortex flows in the dual lattice: counterclockwise (spin up $\uparrow$) and clockwise (spin down $\downarrow$)  vortex currents. It was introduced in Ref.~\cite{PhysRevLett.102.070403} for the hard-core bosons ($U\to\infty$) at half filling and analyzed in detail in Refs.~\cite{PhysRevB.82.134510,pnas}. %We shall briefly review it here. %The structure of the vortex is closely related to the average boson density $n=N/(N_xN_y)$, where $N$ is the total number of the bosons and $N_x (N_y)$ is the size of the system in the $x (y)$ direction. In the low-density limit $n\ll1$, the bosons form a Bose-Einstein condensate (BEC)~\cite{BEC_PhysRevB.65.104519} so that the Hamiltonian can be well understood within the framework of the GP theory~\cite{PhysRevB.82.134510}. The corresponding GP vortex has a large density depletion in the core. In particular, the boson density at the vortex center is zero. In contrast, the structure of the vortex changes considerably when $n$ is close to $\frac{1}{2}$. In this situation, 
%Near half filling, the GP theory is not applicable, since the condensate fraction of the bosons is only about $40\%$~\cite{BEC_PhysRevB.65.104519}. Instead, a semiclassical analysis using the anisotropic nonlinear $\sigma$ model (NLSM) shows that the vortex can be viewed as a meron~\cite{arxiv0701571v1,PhysRevB.82.134510}. It describes the circulation of the superfluid order parameter $n_{\bm{r}}^x+in_{\bm{r}}^y=2\langle a^{\dagger}_{\bm{r}}\rangle=\sqrt{1-(n_{\bm{r}}^z)^2}e^{in\theta(\bm{r})}$, where $n_{\bm{r}}^z=(-1)^{j+k}(2\langle n_{\bm{r}}\rangle-1)$ is the boson CDW in the vortex core, $\theta(\bm{r})$ is the angle between the position vector $\bm{r}=(j,k)a$ and the $x$ axis and $n$ is an integer defining the topological charge of the vortex ($|n|$ is the number of vortices). The CDW decays exponentially from the vortex center $n_{\bm{r}}^z\sim e^{-r/\xi_z}$, where $\xi_z=\sqrt{\frac{V}{2(J-V)}}$ is the decay constant. When $r\gg\xi_z$, $\langle n_{\bm{r}}^z\rangle\to 0$ and the system is in a superfluid phase with the order parameter $n_{\bm{r}}^x+in_{\bm{r}}^y\neq0$. Returning to the original Hamiltonian Eq.~(\ref{Hamiltonian}), the quantum fluctuations of the position of the vortex center becomes important: the vortex will circulate around the lattice sites where the boson charges act as the effective magnetic field, resulting in the clockwise and counterclockwise vortex currents~\cite{pnas}. The two circulating forms of the vortex currents, referred to as the vortex spin~\cite{PhysRevLett.102.070403,PhysRevB.82.134510}, 
The vortex spin corresponds to the doubly degenerate ground states of the Hamiltonian Eq.~(\ref{Hamiltonian}) with a single flux quantum penetrating the system ($N_\Phi=BN_xN_ya^2/\Phi_0=1$).

\emph{The entanglement of vortices}. The entanglement between the two vortex spins was proposed in Ref.~\cite{arxiv0701571v1} for the system with two flux quanta ($N_\Phi=2$). The two vortices are pinned around the two antipodal lattice sites through weakening the bonds connected to them~\cite{arxiv0701571v1,supp}. The bonds weakening can be realized in experiments by engineering the waveform of the laser beams using the quantum gas microscope and the digital micromirror device~\cite{Greiner_Nature,Preiss-phdthesis}. Fig.~\ref{Fig-symmetry-axis} shows the location of the two pinning sites $1,2$ in a $4\times4$ square lattice with periodic boundary conditions (the torus geometry), where the Landau gauge is chosen for the vector potential $\bm{A}$: $A_x=-yBN_x\delta_{x,N_x-1}$, $A_y=xB$. Denote the hopping strength for the bonds involving the sites $1$ and $2$ as $J_{\textrm{pin}}$ and the hopping strength for other bonds as $J$. When $J_{\textrm{pin}}/J>1$, the vortices are repelled by the pining sites and will be located far from them. On the contrary, when $J_{\textrm{pin}}/J<1$, the vortices are attracted by the pining sites and will be located close to them. 

The pinning effect can be verified by plotting the vorticity of the electric current density $\bm j(\bm r)=-\frac{\partial H}{\partial\bm A(\bm r)}$. Fig.~\ref{Fig-vtc} shows $\langle\nabla\times\bm j\rangle$ in the ground state for different pinning strength. It can be seen that the vorticity peaks far from the sites $1,2$ for $J_{\textrm{pin}}/J=1.5$, while it is uniformly distributed for $J_{\textrm{pin}}/J=1$. As $J_{\textrm{pin}}/J$ decreases further to $0.9$, the vorticity around the sites $1,2$ becomes larger than that in the other plaquettes, and finally it peaks around the two pinning sites for $J_{\textrm{pin}}/J=0.6$.

For generic pinning strength, the ground state of the Hamiltonian~(\ref{Hamiltonian}) is nondegenerate and can be labelled by the maximally entangled state of the two vortex spins: $\ket{G}=\frac{1}{\sqrt{2}}(\ket{\uparrow_1\downarrow_2}+\ket{\downarrow_1\uparrow_2})$, where $\uparrow_j$ ($\downarrow_j$) represents the up (down) component of the vortex spin at the site $j$. The state $\ket{\omega_j}$ with $\omega=\uparrow,\downarrow$ can be written as $\ket{\omega_j}=c_{j\omega}^\dagger\ket{\textrm{vac}}$, where $c_{j\omega}^\dagger$ creates the vortex current $\omega_j$ on the vacuum state $\ket{\textrm{vac}}$ with no vortices, $c_{j\omega}^\dagger=\sum_{\bm R}f(|\bm R-\bm r_j|)e^{im_\omega\varphi(\bm R-\bm r_j)}b_{\bm R}^\dagger$. The operator $b_{\bm R}^\dagger$ creates a vortex on the dual lattice site $\bm R$, $f(|\bm R-\bm r_j|)$ is a function obtained by diagonalizing the effective vortex-hopping Hamiltonian~\cite{pnas} and has the maximum at the site $j$ with the position vector $\bm r_j$, $\varphi(\bm R-\bm r_j)$ is the angle between $\bm R-\bm r_j$ and the $x$ axis, and $m_\omega=0,1$ for $\omega=\downarrow,\uparrow$ respectively. An effective Hamiltonian responsible for the state $\ket{G}$ can be constructed~\cite{supp}:
\ba
H_\textrm{s}=f_{xx}(\tau_{1}^x\tau_{2}^x+\tau_{1}^y\tau_{2}^y)+f_{zz}\tau_{1}^z\tau_{2}^z, 
\ea
where $\tau_j^\alpha=\frac{1}{2}\sigma_j^\alpha$ with $\sigma_j^\alpha$ being the Pauli matrices, the $f_{xx}$ term denotes the tunneling between the two vortex spins (the kinetic energy), and the $f_{zz}$ term stems from the Coulomb interactions between vortices~\cite{Ueda_book}. The coefficients $f_{xx},f_{zz}$ are determined by matching the spectrum of $H_\textrm{s}$ with the low-lying levels of $H$ in~(\ref{Hamiltonian}).

It is rather difficult to directly detect the entanglement of vortices through measuring the correlation functions of the bosons that comprise the vortex. This is because the vortex describes the circulation of the superfluid which is macroscopic consisting of a large number of bosons. %For instance, the $z$ component of the vortex spin is represented by the meron density operator $\tau_{\bm{r}}^z\approx\frac{1}{4\pi}\hat{\bm{n}}\cdot D_x\hat{\bm{n}}\times D_y\hat{\bm{n}}$, where $D_\alpha=\partial_\alpha-iqA_\alpha$ and $\hat{\bm{n}}=(n_{\bm{r}}^x,n_{\bm{r}}^y,n_{\bm{r}}^z)$. This operator involves multiple lattice sites in the vortex core. To detect the entanglement of vortices, one needs to measure the correlation function $\langle\tau_1^z\tau_2^z\rangle$ which involves complicated correlation functions of the boson operators. This measurement is rather difficult in general. %Moreover, the $x,y$ components of the vortex spin operator, indispensable for detecting the entanglement of vortices, are difficult to define. 
However, the difficulty can be circumvented by detecting the CDW order alone that competes with the superfluid order. The entanglement of vortices is accompanied by the coherent CDWs in the two vortex cores~\cite{supp}. The state of the latter reduces to a maximally entangled state between the two pinning sites when $V\to 0$:
\ba\label{two-sites-etg}
\ket{\Psi_{12}}=\frac{1}{\sqrt{2}}(\ket{10}+e^{i\phi}\ket{01}),
\ea 
where $\phi=\textrm{Arg}\langle a_1^\dagger a_2\rangle$ is related to the gauge field~\cite{supp}. %This conjecture is in accordance with Eq.~(\ref{GS-vortex-etg}): an excess boson density ($\ket{1}$) acts as an effective magnetic field in the positive $z$ direction, resulting in a counterclockwise vortex current; while a reduced boson density ($\ket{0}$) acts as an effective magnetic field in the negative $z$ direction, resulting in a clockwise vortex current.
The state~(\ref{two-sites-etg}) exhibits strong coherence between the two vortex centers. This is reflected by the equal superposition between $\ket{10}$ and $\ket{01}$ for which the modulus of the correlation function $|\langle a_1^\dagger a_2\rangle|=\frac{1}{2}$ is maximized. 

\begin{figure}[b]
%\begin{center}
%\vspace{10cm}
%\subfigure[]{\label{Fig-1-a}
\includegraphics[width=3.6in]{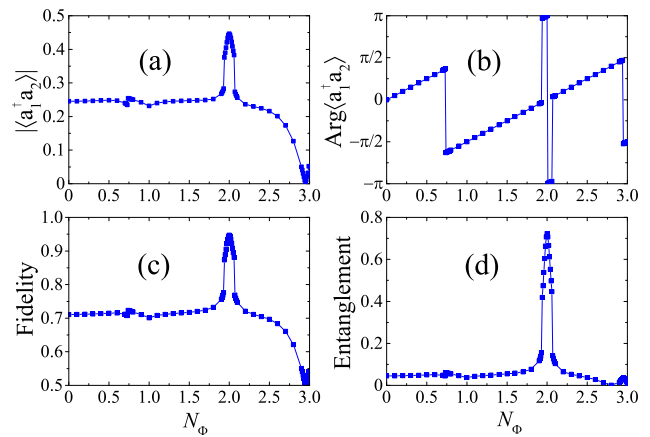}%}
%\vspace{3mm}
%\subfigure[]{\label{Fig-1-b}
%\includegraphics[width=3in]{Fig-1-b}}
%\end{center}
\caption{(Color online). The four quantities are plotted against the total flux quanta $N_\Phi$ in a $4\times4$ square lattice with $8$ bosons: (a) The modulus of $\langle a_1^\dagger a_2\rangle$; (b) The phase of $\langle a_1^\dagger a_2\rangle$~\cite{supp}; (c) The fidelity between the reduced state $\rho_{12}$ of the two vortex centers and  $\ket{\Psi_{12}}$ in Eq.~(\ref{two-sites-etg}), where the phase $\phi$ is set with the values in Fig. 1(b); (d) The entanglement of $\rho_{12}$. The symbols in the curves represent the data points.}\label{Fig-1-JSM-0.6}
\end{figure}

To verify the above analysis, the modulus and phase of $\langle a_1^\dagger a_2\rangle$ versus the total flux quanta $N_\Phi$ are plotted in Figs.~\ref{Fig-1-JSM-0.6}(a) and \ref{Fig-1-JSM-0.6}(b) for a $4\times4$ square lattice with $8$ bosons, using the exact diagonalization techniques~\cite{ED}. Here, $V=0$ and $J_{\textrm{pin}}/J=0.6$. The flux quanta $N_\Phi$ must be integers for a closed surface~\cite{Dirac60}, but we let it vary continuously in order to analyze the data more explicitly. It can be seen in Fig.~\ref{Fig-1-JSM-0.6}(a) that $|\langle a_1^\dagger a_2\rangle|\approx0.25$ for $0<N_\Phi<2$, reflecting the stable superfluid order of the system, while it peaks at $N_\Phi=2$ with the value close to $\frac{1}{2}$ which is consistent with the state~(\ref{two-sites-etg}). Then it decreases with $N_\Phi$, as the superfluid order is generally suppressed by increasing the magnetic flux. Fig.~\ref{Fig-1-JSM-0.6}(c) shows the fidelity~\cite{Nielsen} between the reduced state $\rho_{12}$ of the two vortex centers and  $\ket{\Psi_{12}}$ in Eq.~(\ref{two-sites-etg}), where the phase $\phi$ is set with the values in Fig. 1(b). The fidelity is close to $1$ when $N_\Phi=2$, which clearly corroborates the conjecture. The entanglement of formation~\cite{Wootters_etg} of $\rho_{12}$ has a sharp peak value $\sim0.72$ at $N_\Phi=2$ as shown in Fig.~\ref{Fig-1-JSM-0.6}(d). Later, we shall show that for a stronger pinning strength the entanglement can approach one consistent with Eq.~(\ref{two-sites-etg}).%depends on the pinning strength: a stronger pinning strength results in a larger entanglement which can approach one consistent with Eq.~(\ref{two-sites-etg}).

\emph{Experimental detection.} Conventionally, the bosonic coherence can be detected through observing the interference fringes in the time-of-flight image~\cite{PhysRevLett.95.050404,PhysRevA.72.053606,PhysRevA.79.053623,PhysRevA.84.053613,PhysRevA.87.033614}. However, as the interference fringes involve the contributions from the correlation functions $\langle a_{\bm r}^\dagger a_{\bm r'}\rangle$ of all sites, it is difficult to single out the contribution from the two pinning sites. Here, we propose two methods to determine $\langle a_1^\dagger a_2\rangle$ for the scheme of the optical lattice with $^{87}$Rb atoms.

Firstly, the quantum interference between the two pinning sites can be realized through coupling them using two Raman lasers detuned by $\Delta'$ (the energy offset between the two pinning sites). The effective Hamiltonian $H'=-(\Omega_{\textrm{eff}}a_1^\dagger a_2+\textrm{H.c.})+U\sum_{j}n_j(n_j-1)$, where $\Omega_{\textrm{eff}}$ is the effective Rabi frequency of the field for coupling the two sites, $U$ is the on-site interaction strength, and $n_j=a_j^\dagger a_j$. We assume that the two Raman lasers do not couple the pinning sites to other sites. This requires applying a nonlinear magnetic field gradient to the lattice such that $\Delta'$ is different from the energy offset between one of the pinning sites and any other site. The magnetic field gradient also inhibits the direct tunneling between the neighboring sites. In this way, $\langle a_1^\dagger a_1\rangle_t$=$\textrm{Tr}(a_1^\dagger a_1e^{-iH't}\rho_{12}e^{iH't})$=$\frac{1}{2}-r\sin(2|\Omega_{\textrm{eff}}|t)\sin(\phi'+\theta')$ satisfying $\langle a_1^\dagger a_1\rangle$=$\frac{1}{2}$ at $t=0$ for half filling ($\langle a_1^\dagger a_2\rangle$$\equiv$$re^{i\phi'}$, $\theta'$=$\textrm{Arg}(\Omega_{\textrm{eff}})$). %$
Thus, $r$ and $\phi'$ are determined through measuring $\langle a_1^\dagger a_1\rangle_t$ achieved by a quantum gas microscope~\cite{Greiner_Nature,Bakr547}.

Secondly, the density matrix $\rho_{12}$ of the two pinning sites can be constructed by tomography. In the hard-core limit, $\rho_{12}$ has a simple form: $\rho_{12}=\sum_{i,j=0,1}x_{i,j}\ket{ij}\bra{ij}+(y\ket{01}\bra{10}+\textrm{H.c.})$. The diagonal matrix elements $x_{i,j}$ can be determined by using the quantum gas microscope which measures the even-odd parity of the atoms at individual sites~\cite{Greiner_Nature,Bakr547}, %For instance, $x_{00}$ is equal to the probability of the event that both the pinning sites have even parity for the atoms. This probability can be deduced through repeatedly preparing the system in the same ground state, imaging the parity of the atom in each time, and calculating the fraction of the above event ($=x_{00}$). 
while $|y|$ can be determined by measuring the purity of $\rho_{12}$: $\textrm{Tr}(\rho_{12}^2)=|y|^2+\sum_{i,j=0,1}x_{i,j}^2$, which is realized through interfering the two identical copies of the system~\cite{Greiner_etg}. The phase of $y$ is undetermined, but this does not affect the coherence ($|\langle a_1^\dagger a_2\rangle|=|y|$).

\begin{figure}[t]
%\begin{center}
%\vspace{10cm}
%\subfigure[]{\label{Fig-1-a}
\includegraphics[width=3in]{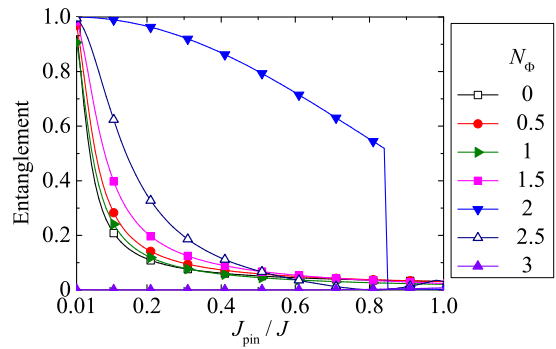}%}
%\vspace{3mm}
%\subfigure[]{\label{Fig-1-b}
%\includegraphics[width=3in]{Fig-1-b}}
%\end{center}
\caption{(Color online). The entanglement of the reduced state $\rho_{12}$ for the two vortex centers is plotted against the ratio $J_{\textrm{pin}}/J$ for different values of the flux quanta $N_{\Phi}$. Here, $J_{\textrm{pin}}$ is the hopping strength for the bonds involving the sites $1$ and $2$, and $J$ is the hopping strength for other bonds.}\label{Fig-etg-Jpin}
\end{figure}

The pinning of the two entangled vortex spins plays a crucial role in establishing the coherence of the two sites as in Eq.~(\ref{two-sites-etg}). Fig.~\ref{Fig-etg-Jpin} shows the entanglement of formation of $\rho_{12}$ versus $J_{\textrm{pin}}/J$ for different values of the flux quanta $N_{\Phi}$~\cite{Jpin_deg}. It can be seen that when $J_{\textrm{pin}}/J$ is small, the entanglement for most cases is close to $1$ except for $N_{\Phi}=3$. This entanglement is a perturbative effect: the coherence between the two sites is established through their coupling to the many-body states of the Hamiltonian with $J_{\textrm{pin}}=0$, analogous to the superexchange effect~\cite{supp}. The entanglement decreases rapidly with $J_{\textrm{pin}}/J$ for generic values of $N_{\Phi}\neq2$. In contrast, when $N_{\Phi}=2$, the entanglement decreases slowly up to a critical value of $J_{\textrm{pin}}/J\approx0.85$ where the entanglement suddenly drops to a value close to zero ($\approx0.003$). These results indicate that the two entangled vortex spins, when pinned to the two appropriate lattices sites, can protect the coherence established in the two sites for a wide range of pinning strength. When $J_{\textrm{pin}}/J\gtrsim0.85$, the vortices delocalize and the entanglement becomes negligible.

The strong bosonic coherence between the two vortex centers for $N_{\Phi}=2$ can be viewed as a qubit~\cite{parafermion} stored in the ground state of the system as in Eq.~(\ref{two-sites-etg}), with the basis states encoded as $\ket{\tilde{0}}=\ket{10}$ and $\ket{\tilde{1}}=\ket{01}$. By tuning $J_{\textrm{pin}}/J$ to be lower than $0.85$, the coherent information (the entanglement of vortices) in the ground state is retrieved in the form of this qubit, which can be harnessed to produce entangled qubits~\cite{supp} for use in the teleportation-based quantum computation (TQC)~\cite{PhysRevA.70.060302}. TQC applies to the situations where quantum gates between remote qubits are needed. We notice that the retrieving process for generic values of $N_{\Phi}\neq2$ is also possible, but it requires a rather low value of $J_{\textrm{pin}}/J$, as shown in Fig.~\ref{Fig-etg-Jpin}. Therefore, our proposal provides a feasible scheme of quantum memory for storing qubits useful in quantum computation.

\emph{Conclusion}. We have shown that the entanglement of vortices, resulting from the charge-vortex duality, can be detected through the interference of the bosons in the two vortex centers achieved using the Raman coupling and the quantum gas microscope. The coherent bosons in the two vortex centers can be viewed as a qubit stored in the ground state of the system. The coherence of this qubit can be harnessed to produce entangled qubits useful in quantum computation. It is expected that the present work can be generalized to higher dimensionality~\cite{PhysRevLett.111.115303,Baier201,PhysRevA.94.063641,Cirac_NJP} and also to other systems with multiple orders, including the spin-orbit coupled bosons with a non-Abelian gauge field where the magnetic order coexists with the superfluid order~\cite{PhysRevLett.109.085302} and the fermionic superfluids~\cite{PhysRevA.81.031607,Huang_Fermi,PhysRevA.95.043615}. %The latter entanglement indicates that there is strong coherence between the two vortex centers, in contrast to the core depletion of the vortex in the Bose-Einstein condensate. The bosonic entanglement in the vortex centers can be viewed as a qubit that is stored in the ground state of the system. The coherence of the qubit is robust against the varying of the pinning strength for the vortices to a wide range. Therefore, our proposal provides a scheme of quantum memory for storing qubits. %Future work can be pursued on the scenario with more flux quanta, where the entanglement of multi-vortices is created. It is expected that the coherence of the two bosons studied in the present work will extends to more bosons, establishing the multi-boson entanglement. An implication is that the entanglement among the bosons in a subsystem of the multi-boson state should be very weak, as shown in Fig.~\ref{Fig-etg-Jpin} with $N_\Phi=3$.

%\section*{Acknowledgements}
We acknowledge Ning-Hua Tong for helpful discussions. This work was supported by the National Key R\&D Program of China under grants Nos. 2016YFA0301500, NSFC under grants Nos. 11434015, 61227902, 61835013, SPRPCAS under grants No. XDB01020300, XDB21030300.

\section{\Large\textbf Supplemental Material}

\section{The vortex pinning effect}
A vortex can be pinned to some lattice site through decreasing the strength of the tunneling between this site and its neighbors. This effect is understood by considering the change of the energy in the pinning process. The classical Hamiltonian of the two-dimensional Bose-Hubbard model is~\cite{Sachdev_book}
\ba
H^{cl}&=&H_0^{cl}+H^{cl}_\delta,\label{H_all}\\
H^{cl}_0&=&-2J\sum_{\langle\bm{rr'}\rangle}\sqrt{\rho_{\bm r}\rho_{\bm r'}}\cos(\phi_{\bm r}-\phi_{\bm r'}-A_{\bm{rr'}})\nonumber\\
&&+U\sum_{\bm r}\rho_{\bm r}^2+V\sum_{\langle\bm{rr'}\rangle}\rho_{\bm r}\rho_{\bm r'}-\mu\sum_{\bm r}\rho_{\bm r},\label{H_0}\\
H^{cl}_\delta&=&-2J_\delta\sum_{\langle\bm{r_0r'}\rangle}\sqrt{\rho_{\bm r_0}\rho_{\bm r'}}\cos(\phi_{\bm r_0}-\phi_{\bm r'}-A_{\bm{r_0r'}}),\label{H_delta}
\ea
where $H^{cl}_0$ is the original Hamiltonian with uniform tunneling strength $J$, and $H^{cl}_\delta$ describes the energy change when $J$ is changed to $J_\delta$ for a particular site $\bm r_0$ ($J_\delta=J_{\textrm{pin}}-J$). The parameters $\rho_{\bm r}$ and $\phi_{\bm r}$ are the density and phase of the coherent state $\ket{\beta_{\bm r}}$ at the site $\bm r$: $a_{\bm r}\ket{\beta_{\bm r}}=\beta_{\bm r}\ket{\beta_{\bm r}}$, $\beta_{\bm r}=\sqrt{\rho_{\bm r}}e^{i\phi_{\bm r}}$.

A general vortex configuration is defined as $\beta_{\bm r}=\sqrt{\rho_{\bm r}}\exp[i\sum_{k=1}^{N}n_k\theta(\bm r-\bm r_k)]$. It represents $N$ vortices centered at $\bm r_k$ with the topological charge $n_k$ (an integer), $k=1,2,\cdots,N$. The parameter $\theta(\bm r-\bm r_k)$ is the angle between the vector $\bm r-\bm r_k$ and the $x$ axis. The simplest case is a single vortex centred at $\bm r_1$ ($N=1$ and $n_1=1$).  If the vortex center $\bm r_1$ is far from the pinning center $\bm r_0$, we have $\phi_{\bm r_0}-\phi_{\bm r'}=\theta(\bm r_0-\bm r_1)-\theta(\bm r'-\bm r_1)\approx 0$ in Eq.~(\ref{H_delta}) and $H^{cl}_\delta\approx-2J_\delta\sum_{\langle\bm{r_0r'}\rangle}\sqrt{\rho_{\bm r_0}\rho_{\bm r'}}$, where we have assumed that the magnetic flux per unit cell is very small so that $A_{\bm{r_0r'}}\approx0$ (this is true if the number of the lattice sites is large and the total magnetic flux is small). When $\bm r_1$ approaches $\bm r_0$, $\phi_{\bm r_0}-\phi_{\bm r'}$ increases. Therefore, $H^{cl}_\delta$ decreases if $J_\delta<0$, whereas it increases if $J_\delta>0$. This implies that the pinning center attracts the vortex if $J_{\textrm{pin}}<J$, but it repels the vortex if $J_{\textrm{pin}}>J$. $H^{cl}_\delta$ is minimized when $\bm r_1$ is located in between $\bm r_0$ and one of its neighbors $\bm r'$ for which $\cos(\phi_{\bm r_0}-\phi_{\bm r'}-A_{\bm{r_0r'}})\approx-1$. In this case, $H^{cl}_\delta\approx0$ for a square lattice where $\bm r_0$ has four neighbors. If the number of the lattice sites is small, $A_{\bm{r_0r'}}$ in Eq.~(\ref{H_delta}) may not be negligible. A detailed analysis of each term in $H^{cl}_\delta$ shows that the conclusion regarding the vortex pining effect still holds.

The scenario with multiple vortices and multiple pinning centers can be analyzed in a similar way. In the hard-core limit, it is more convenient to use the spin half coherent state path integral to obtain the classical Hamiltonian~\cite{Sachdev_book,PhysRevB.82.134510}. The discussion on these cases are omitted.

We notice that as shown in Fig.~2 of the main text, when $J_{\textrm{pin}}/J=1.5$, the vorticity peaks around the two sites $3,4$ labeled in Fig.~1(b), which can be viewed as a vortex anti-pinning effect. Interestingly, this effect also results in a strong coherence between the sites $3$ and $4$ which is verified numerically. However, this is the special case for the lattice of a small size ($4\times4$), since the unpinned vortex pair is situated either around the sites $1,2$ or around the sites $3,4$ to minimize the classical energy~\cite{arxiv0701571v1}. For the lattice of a large size, there will be more alternative sites that the vortex pair can move for when it is repelled by the pinning sites $1,2$.

\section{Estimation of the background vorticity}
In Fig.~2 of the main text, the the vorticity of the electric current density $\langle\nabla\times\bm j\rangle$ is subtracted by the background vorticity ($\approx-0.43\frac{eJ}{\hbar a^2}$ per plaquette of the lattice). The background vorticity is a lattice effect. %First, we notice that the total magnetic flux of a closed surface is always zero because $\nabla\cdot\bm B=0$. This would imply that the lattice with the torus geometry cannot be penetrated by nonzero flux quanta. However, the lattice is a discretized surface. The flux quanta of each plaquette of the lattice has the uncertainty of an arbitrary integer. This is because the Hamiltonian in Eq.~(1) of the main text remains the same if $A_{\bm{rr'}}$ for any $\langle\bm{rr'}\rangle$ is changed by multiples of $2\pi$, but the latter will induce a change of the flux quanta of the plaquettes that involve $A_{\bm{rr'}}$ by an amount of some integers. We shall show that this phenomenon can be utilized to create effectively a nonzero flux quantum for the lattice with the torus geometry. 
Let us consider Fig.~1(b) of the main text for concreteness. The magnetic flux of each plaquette is $\Phi/(N_xN_y)=Ba^2$, where $\Phi$ is the total flux quanta, $N_x (N_y)$ is the size of the system in the $x (y)$ direction, $B$ is the magnetic field and $a$ is the lattice constant. The only exception is for the plaquette in the upper right corner whose magnetic flux is $\oint_{\bm{r}}^{\bm{r'}}\bm{A}\cdot d\bm{l}=Ba^2-\Phi$. In this way, the total magnetic flux of the lattice is $Ba^2(N_xN_y-1)+(Ba^2-\Phi)=0$ as required by
$\nabla\cdot\bm B=0$.

To observe the physical effect of the lattice with the magnetic flux $Ba^2$ per plaquette, the additional flux $-\Phi$ in the upper right corner need be subtracted in an appropriate way. If $N_\Phi=\Phi/\Phi_0$ is an integer, the flux $-\Phi$ can be moved to the neighboring plaquette through increasing the vector potential of any one edge of the upper right corner by $\Phi/a$, which does not change the Hamiltonian. In fact, 
the flux $-\Phi$ can be moved further to any plaquette by a similar operation on the vector potential. Therefore, one can envision that the lattice is penetrated by a uniform magnetic flux $Ba^2$ per plaquette, plus the flux $-\Phi$ for every single plaquette with equal probability $p=1/(N_xN_y)$. The vorticity $\langle\nabla\times\bm j\rangle$ corresponding to the latter flux is the background vorticity. For the flux $-\Phi$ penetrate a single plaquette with the probability $p$, $\langle\nabla\times\bm j\rangle\approx p\oint\langle\bm j\rangle\cdot d\bm l/a^2=-pne\frac{\hbar}{M}\oint\nabla\phi\cdot d\bm l/a^2=-pne\frac{h\Phi}{M\Phi_0a^2}$, where we have assumed that the bosons are in a superfluid phase with $\langle\bm j\rangle=-ne\frac{\hbar}{M}\nabla\phi$. Here, $n$ ($M$) is the density (mass) of the boson, and $\phi$ is the phase of the marcroscopic wavefunction. $M\approx\frac{\hbar^2}{Ja^2}$ by considering the low-energy physics of the Hamiltonian. Thus, $\langle\nabla\times\bm j\rangle\approx-2\pi pne\frac{h\Phi J}{\Phi_0\hbar}$. For the $4\times4$ lattice with two flux quanta at half filling, $n=\frac{1}{2a^2}$, $p=\frac{1}{16}$, $\Phi=2\Phi_0$. $\langle\nabla\times\bm j\rangle\approx-\frac{\pi}{8}\frac{eJ}{\hbar a^2}\approx-0.4\frac{eJ}{\hbar a^2}$ which is close to $-0.43\frac{eJ}{\hbar a^2}$ in the numerical calculation. The latter is obtained by observing that the vortices are completely attracted to the pinning sites for $J_{\textrm{pin}}/J\ll1$ so that the vorticity in other sites is only composed of the background vorticity.

\section{The entangled state of vortices}
The entangled state of vortices $\frac{1}{\sqrt 2}(\ket{\uparrow_1\downarrow_2}+\ket{\downarrow_1\uparrow_2})$ was proposed in Ref.~\cite{arxiv0701571v1}. Here, we derive it in an alternative way based on the interactions between vortex spins. In Eq.~(32) of Ref.~\cite{arxiv0701571v1}, an effective Hamiltonian is used to describe the low-energy physics for the two unpinned vortices. In the presence of pinning, the $z$-component of the total orbital angular momentum of the vortices $L_{\textrm{rel}}^z$ (relative to the vorticity center) is no longer conserved, since the pining potential is not rotationally invariant. Assume the pinning strength is so strong that the vortices are localized around the pinning centers with $L_{\textrm{rel}}^z$ frozen. In this case, only the spin angular momentum of the vortices need be considered. A general Hamiltonian describing the two interacting vortex spins is $H_\textrm{s}=\sum_{\alpha,\beta}f_{\alpha\beta}\tau_{1}^\alpha\tau_{2}^\beta+\bm h_1\cdot\bm\tau_1+\bm h_2\cdot\bm\tau_2$, where $\alpha,\beta=x,y,z$, $f_{\alpha\beta}$ is the coefficient of the spin-spin interaction terms, $\bm\tau_j=(\tau_j^x,\tau_j^y,\tau_j^z)$ are the three components of the vortex spin $j$ ($\tau_j^\alpha=\frac{1}{2}\sigma_j^\alpha$ with $\sigma_j^\alpha$ being the Pauli matrices), and $\bm h_j$ is the effective magnetic field. As $\tau_{\textrm{tol}}^z=\tau_{1}^z+\tau_{2}^z$ is conserved (commutes with $H_\textrm{s}$), $H_\textrm{s}$ is reduced to
\ba
H_\textrm{s}=f_{xx}(\tau_{1}^x\tau_{2}^x+\tau_{1}^y\tau_{2}^y)+f_{zz}\tau_{1}^z\tau_{2}^z+h_1^z\tau_{1}^z+h_2^z\tau_{2}^z. 
\ea
The $f_{xx}$ term denotes the tunneling between the two vortex spins (the kinetic energy), while the $f_{zz}$ term originates from the Coulomb interactions between vortices.

Since the boson charges act as the effective magnetic fields, we have $h_j^z\propto\langle n_j\rangle-\frac{1}{2}$ in the mean-field approximation, where $n_j$ is the number operator for the boson in the site $j$. For the half filling, $h_j^z\approx0$. By fitting $H_\textrm{s}$ with the exact low-energy spectrum of the Bose-Hubbard Hamiltonian with e.g. $J_{\textrm{pin}}=0.6$ (set $J=1$), we find two solutions: $f_{xx}=\pm0.06309$, $f_{zz}=0.2236$ up to a constant ($-13.28$) for $H_\textrm{s}$. Correspondingly, the ground state is $\frac{1}{\sqrt 2}(\ket{\uparrow_1\downarrow_2}\mp\ket{\downarrow_1\uparrow_2})$. To determine which solution is physically correct, we demand that the statistics of the vortices be consistent for all the eigenstates of the Hamiltonian. This implies that the three eigenstates of $H_s$: $\ket{\uparrow_1\uparrow_2}$, $\ket{\downarrow_1\downarrow_2}$, $\frac{1}{\sqrt 2}(\ket{\uparrow_1\downarrow_2}+\ket{\downarrow_1\uparrow_2})$, symmetric under the spin exchange, should have the same exchange property for the spatial part of the wavefunction. This is known by examining the value of $\Pi^z$ defined in Ref.~\cite{arxiv0701571v1}, and we find that $f_{xx}=-0.06309J$ and the ground state is $\frac{1}{\sqrt 2}(\ket{\uparrow_1\downarrow_2}+\ket{\downarrow_1\uparrow_2})$.

\section{The coherent charge-density waves}
Let us assume that the dual lattice for the vortices is cut into two sublattices such that each sublattice contains one vortex pinned in its center. The ground state of the sublattice $j$ is doubly degenerate: $\ket{G_j}=\alpha_j\ket{e_j}\ket{\uparrow_j}+\beta_j\ket{d_j}\ket{\downarrow_j}$, where $\alpha_j,\beta_j$ are the coefficients, and $\ket{\uparrow_j},\ket{\downarrow_j}$ are the spin states of the vortex $j$. The state $\ket{e_j}$ ($\ket{d_j}$) is the state of the bosons in the original lattice with excess (depleted) density at the pinning site $j$~\cite{pnas}, representing the charge-density wave (CDW) that breaks the particle-hole symmetry (PHS). The state $\ket{G_j}$ has the physical meaning that the excess (depleted) boson density at the site $j$ acts as a positive (negative) magnetic field, which induces the counterclockwise (clockwise) vortex current.

The ground state of the whole lattice is $\ket{G}=\ket{G_1}\ket{G_2}=\alpha_1\alpha_2\ket{\uparrow_1\uparrow_2}\ket{e_1e_2}+\beta_1\beta_2\ket{\downarrow_1\downarrow_2}\ket{d_1d_2}+\ket{\Psi_\pm}(\alpha_2\beta_1\ket{e_1d_2}\pm\alpha_1\beta_2\ket{d_1e_2})/\sqrt{2}$, where $\ket{\Psi_\pm}=(\ket{\uparrow_1\downarrow_2\pm\downarrow_1\uparrow_2})/\sqrt{2}$. Then, reconnect the two sublattices which induces an effective spin-spin interactions between the two vortices making the entangled state $\ket{\Psi_+}$ energetically favorable as shown in the previous section. The system will evolve and come to equilibrium with the $\ket{\Psi_+}$ component of $\ket{G}$ as the new ground state. Correspondingly, the two CDWs are in the coherent superposition state $\alpha_2\beta_1\ket{e_1d_2}+\alpha_1\beta_2\ket{d_1e_2}$ up to normalization. This state is entangled in general. It has free coefficients which are determined by requiring its symmetries (e.g. PHS) to be consistent with those of the Hamiltonian for the nondegenerate ground state.

The whole process is analogous to the entanglement swapping~\cite{PhysRevLett.71.4287}, where the entanglement in the vortex-CDW pairs is transferred to the two CDWs through the Bell measurement on the two vortices (which creates the vortex entanglement). Consider the limit $V\to 0$ for which the decay constant $\xi_z=\sqrt{\frac{V}{2(J-V)}}$ of the CDW goes to zero~\cite{PhysRevB.82.134510}. In this situation, the CDW ($\sim e^{-r/\xi_z}$) is nonzero only at the vortex center, and the state of the coherent CDWs reduces to a maximally entangled state between the two pinning sites (Eq.~(2) of the main text). The particle-hole symmetry is restored for this state ($\langle a^\dagger_ja_j\rangle=\frac{1}{2}$), since the ground state is nondegenerate. Notice that the vortex entanglement is a sufficient but not necessary condition for the two pinning sites to be entangled, as shown in the next section.

\section{The phase of $\langle a_1^\dagger a_2\rangle$}
The phase of $\langle a_1^\dagger a_2\rangle$ in Fig.~3(b) of the main text can be understood by considering the symmetry of the Hamiltonian. We notice that when $N_x=N_y$, the Hamiltonian commutes with the symmetry operator $\mathcal{M}=UCR_{12}$, where $R_{12}$ is the reflection through the axis connecting the sites $1$ and $2$ as shown in Fig.~1(b) of the main text, $C$ is the charge conjugation $C=\exp[i\pi\sum_{\bm{r}}(a_{\bm{r}}+a_{\bm{r}}^{\dagger})]$, and the unitary operator $U$ is a gauge transformation: $U=\exp[i\sum_{\bm{r}}\chi^z(\bm{r})(a^{\dagger}_{\bm{r}}a_{\bm{r}}-\frac{1}{2})]$ with $\chi^z(\bm{r})=\frac{2\pi}{\Phi_0}\int^{\bm{r}}d\bm{r'}[\bm{A}(\bm{r'})+\tilde{\bm{A}}(\bm{r'})]$. Here $\tilde{\bm{A}}$ is the vector potential obtained through performing the reflection operation on the original $\bm{A}$.
%We notice that when $N_\Phi$ is an integer, %the reduced state $\rho_{12}$ of the two lattice sites can be written as $\rho_{12}=\sum_{i,j=0,1}x_{i,j}\ket{ij}\bra{ij}+y\ket{01}\bra{10}+y^*\ket{10}\bra{01}$ and $\langle a_1^\dagger a_2\rangle=y$. the Hamiltonian commutes with the symmetry operator $\Pi_{\textrm{V}}^z=UP_{\textrm{V}}^z$, where $P_{\textrm{V}}^z$ is the inversion around the vorticity center $R_{\textrm{V}}$~\cite{PhysRevB.82.134510}. The unitary operator $U$ is a gauge transformation: $U=\exp[i\sum_{\bm{r}}\chi^z(\bm{r})a^{\dagger}_{\bm{r}}a_{\bm{r}}]$ with $\chi^z(\bm{r})=\frac{2\pi}{\Phi_0}\int^{\bm{r}}d\bm{r'}[\bm{A}(\bm{r'})-\tilde{\bm{A}}(\bm{r'})]$. Here $\tilde{\bm{A}}$ is the vector potential obtained through performing an inversion operation on the original $\bm{A}$. 
We have $\langle a_1^\dagger a_2\rangle=\langle\mathcal{M}^\dagger a_1^\dagger a_2\mathcal{M}\rangle=\langle a_1 a_2^\dagger e^{i\pi N_{\Phi}}\rangle=\langle a_1^\dagger a_2\rangle^*e^{i\pi N_{\Phi}}$. Thus, 
\ba\label{arg_a1da2}
\textrm{Arg}\langle a_1^\dagger a_2\rangle=\frac{\pi}{2}N_{\Phi}+m\pi,
\ea
where $m$ is an integer.

Eq.~(\ref{arg_a1da2}) is consistent with Fig.~3(b) of the main text: the slope of the curve is $\frac{\pi}{2}$ with jumps $\pm\pi$ ($\Delta m=\pm1$) at $N_\Phi\approx0.73,1.93,2.06,2.94$. These jumps are attributed mainly to the abrupt change of the ground state structure, as it is numerically verified that the overlap of the ground states with neighboring $N_\Phi$'s at these jumps drops to nearly zero and the energy gap closes up to a small value ($<0.005$) due to the finite-size effect (except $N_\Phi=2.94$ which has a vanishing $|\langle a_1^\dagger a_2\rangle|$).

\section{Perturbation theory}
\begin{figure*}[ht]
%\begin{center}
%\vspace{10cm}
%\subfigure[]{\label{Fig-1-a}
\includegraphics[width=6in]{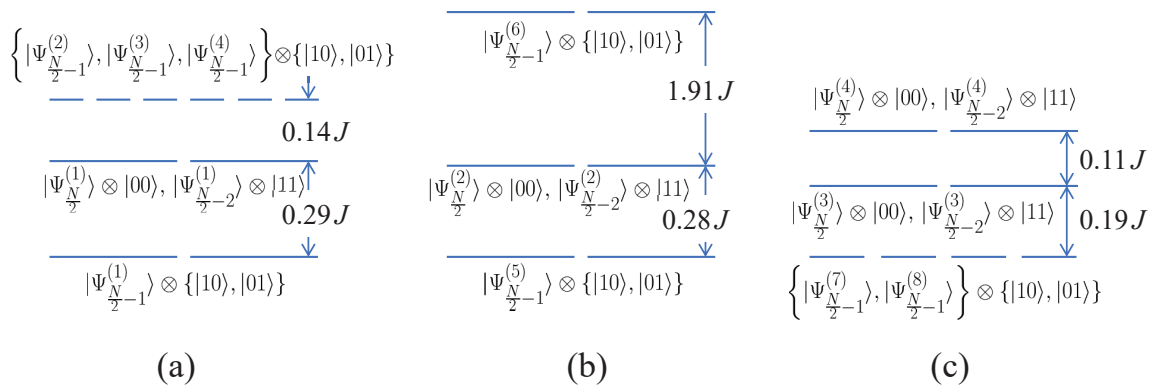}%}
%\vspace{3mm}
%\subfigure[]{\label{Fig-1-b}
%\includegraphics[width=3in]{Fig-1-b}}
%\end{center}
\caption{(Color online). The three low-lying energy levels of the Hamiltonian in Eq.~(\ref{H-perturb-all}) with $J_{\textrm{pin}}=0$ for (a) $N_\Phi=0$, (b) $N_\Phi=0.5$, and (c) $N_\Phi=1$. The state $\ket{\Psi_m^{(n)}}$ denotes an eigenstate of $H_0$ in Eq.~(\ref{H-perturb-all}) where the total number of bosons is $m$ and $n$ is a label to distinguish the different states with the same $m$. The states $\ket{jk}$ ($j,k=0,1$) are the number states of the two pinning sites. The low-lying energy levels for $N_\Phi=1.5,2,2.5$ are similar to (b) with slightly different energy gaps. The low-lying energy levels for $N_\Phi=3$ is similar to (c) with the energy gaps changed from $(0.19J,0.11J)$ to $(0.10J,0.21J)$.}\label{Fig-perturb}
\end{figure*}

In Fig.~4 of the main text, the entanglement for most of the $N_\Phi$'s is close to one when $J_{\textrm{pin}}/J\ll1$. This result can be understood by using the perturbation theory. The Hamiltonian can be written as
\ba
H&=&H_0+V,\label{H-perturb-all}\\
V&=&-J_{\textrm{pin}}\sum_{\langle\bm{r_0r'}\rangle}(e^{iA_{\bm{r_0r'}}}a^{\dagger}_{\bm{r_0}}a_{\bm{r'}}+\textrm{H.c.}),
\ea
where $H_0$ is the Hamiltonian in Eq.(1) of the main text with $J_{\textrm{pin}}=0$, and $V$ is the perturbation for the two pinning sites with $\bm{r_0}=1,2$. The eigenstates of $H$ for $J_{\textrm{pin}}=0$ are obtained by diagonalizing $H_0$ only. We find that for the seven values of $N_\Phi$ in Fig.~4 of the main text, the structure of the low-energy eigenstates of $H$ for $J_{\textrm{pin}}=0$ and half-filling can be classified into three groups as shown in Fig.~\ref{Fig-perturb}. It can be seen that the ground states are doubly degenerate for generic values of $N_\Phi\neq1,3$, where both the two pinning sites and the rest of the system are half-filled.

Let us focus on the case with $N_\Phi=0.5$ in Fig.~\ref{Fig-perturb} (b). When $J_{\textrm{pin}}$ is turned on, the doubly degenerate ground levels will split, which can be analyzed using the degenerate perturbation theory. The effect Hamiltonian in the doubly degenerate ground-state subspace $D$ is
\ba\label{H_eff}
H_{\textrm{eff}}=-\sum_{k\notin D}\frac{V\ket{k^{(0)}}\bra{k^{(0)}}V}{E^{(0)}_k-E^{(0)}_D}=-\bm R\cdot\bm\sigma,
\ea
where $E^{(0)}_D$ is the zeroth-order ground-state energy of $H$, and $\ket{k^{(0)}}$'s are the zeroth-order eigenstates of $H$ with eigen-energies $E^{(0)}_k$. The vector $\bm R=(R_0,R_x,R_y,R_z)$ are the four coefficients and $\bm\sigma=(\sigma_0,\sigma_x,\sigma_y,\sigma_z)$ are the four $2\times2$ matrices ($\sigma_0$ is the identity matrix and the other matrices are the three Pauli matrices). The basis vectors of the matrices are $\ket{\tilde{1}}\equiv\ket{\Psi_\frac{N}{2}}\ket{10}$ and $\ket{\tilde{0}}\equiv\ket{\Psi_\frac{N}{2}}\ket{01}$.

When $J_{\textrm{pin}}/J\ll1$, it is sufficient to include only the first excited states in Eq.~(\ref{H_eff}). For instance, when $J_{\textrm{pin}}/J=10^{-2},N_\Phi=0.5$, the result is $\bm R\approx(5.591,3.866,3.198,0)\times10^{-4}J$. The ground state is $\ket{G}=\ket{\Psi_\frac{N}{2}}(\cos\frac{\theta}{2}\ket{10}+e^{i\phi}\sin\frac{\theta}{2}\ket{01})$ written in the Bloch sphere form, where $\theta=\textrm{ArcCos}(R_z/R_1)=\pi/2$, $\phi=\textrm{ArcTan}(R_y/R_x)=0.69$, ($R_1=\sqrt{R_x^2+R_y^2+R_z^2}$), close to the result from the exact diagonalization ($\theta=1.57,\phi=0.78$). %The energy gap is $2R_1\equiv2\sqrt{R_x^2+R_y^2+R_z^2}\approx0.00358J$ for $J_{\textrm{pin}}/J=10^{-2}$, which is close to the result from the exact diagonalization ($\approx0.00268J$). 
It can be seen that the ground state is a maximally entangled state between the two pinning sites. The coherence between $\ket{10}$ and $\ket{01}$ is established through their coupling to the excited states $\ket{k^{(0)}}$ as in Eq.~(\ref{H_eff}) in which the boson in one pinning site tunnels to the intermediate sites ($\ket{\Psi_{\frac{N}{2}-1}}$ changes to $\ket{\Psi_\frac{N}{2}}$), and then further to the other pinning site. This is analogous to the superexchange effect~\cite{Auerbach_book}.

Interestingly, although the low-lying energy levels for $N_\Phi=2$ is similar to Fig.~\ref{Fig-perturb}(b) with the energy gaps changed slightly from $(0.28J,1.91J)$ to $(0.27J,1.82J)$, a similar perturbation calculation cannot lift the ground-state degeneracy. This implies that the coherence between the two pinning sites for $N_\Phi=2$ is a high-order perturbation effect consistent with the analysis in the previous section, as the latter is based on the vortex-vortex interactions which are expected to involve high energy levels of $H_0$.

The perturbation calculation for Figs.~\ref{Fig-perturb}(a) and \ref{Fig-perturb}(c) can be performed in a similar way but more excited states need be included, which is omitted here. We notice that the ground-state degeneracy is not completely lifted for Fig.~\ref{Fig-perturb}(c), and the ground state is not in the form of Eq.~(2) of the main text for $N_{\Phi}=3$. This exceptional case is due to the fact that the coherence is established among the three sites corresponding to three vortices; any two sites have only negligible entanglement in the three-site entangled state. 

\section{Entangled qubits}
The coherence of the bosons in the two vortex centers can be harnessed to produce entangled qubits. Assume that we have two copies of the system where the pinning sites are denotes as $(1,2)$ and $(1',2')$ respectively. The quantum state of the four sites is close to the two copies of Eq. (2) of the main text: $\ket{\Psi}=\frac{1}{\sqrt{2}}(\ket{10}-\ket{01})\otimes\frac{1}{\sqrt{2}}(\ket{10}-\ket{01})=\frac{1}{2}(\ket{1100}+\ket{0011}-\ket{1001}-\ket{0110})$, where the site sequence of the last expression is changed to $(11'22')$ and the phase $\phi=\pi$ as shown in Fig.~3(b) of the main text.

An even-odd parity measurement of the two sites $11'$ (or $22'$) can be realized by coupling them dispersively with the coherent photons~\cite{PhysRevA.69.062320,PhysRevLett.110.243604,srep_Dai}: $\ket{\Psi}\to\ket{\Psi'}=\frac{1}{2}[(\ket{1100}+\ket{0011})\ket{\alpha}-(\ket{1001}+\ket{0110})\ket{-\alpha}]$, where $\ket{\pm\alpha}$ are the coherent states of the photons and $|\langle\alpha|-\alpha\rangle|=e^{-2|\alpha|^2}\ll1$ for large $\alpha$. The two coherent states can be measured through the homodyne detection with parametric amplifiers and mixers~\cite{PhysRevA.69.062320}. For each measurement result, $\ket{\Psi'}$ collapses to either $\frac{1}{\sqrt{2}}(\ket{1100}+\ket{0011})$ or $\frac{1}{\sqrt{2}}(\ket{1001}+\ket{0110})$ for the four sites. Both states are the maximally entangled state between the qubit $(11')$ and the qubit $(22')$.
 
\section{Finite on-site and nearest-neighbor interactions}

We would like to remark that the charge-conjugation (particle-hole) symmetry of the Hamiltonian is crucial for detecting the vortex entanglement. This is because the vortex spin description is exact only when such symmetry is present~\cite{PhysRevB.82.134510}. This symmetry requires the system to be at half filling and the on-site interaction $U$ to be infinite (the hard-core limit). Breaking the symmetry through changing the filling or decreasing $U$ will induce the energy splitting of the vortex spin states~\cite{pnas} and potentially the spin-spin interactions, which could change the ground state substantially. We numerically verified that at fillings other than one half, $|\langle a_1^\dagger a_2\rangle|$ drops drastically and is less than $0.01$. The result is similar for the finite-$U$ situation. When $U/J\leq10$, $|\langle a_1^\dagger a_2\rangle|<0.01$. This can be interpreted by the GP theory, since for small $U/J$, the system is a BEC and the corresponding GP vortex has a large density depletion in the core. When $10<U/J\leq19.2$, $|\langle a_1^\dagger a_2\rangle|<0.0155$. Further increasing $U/J$ results in a drastic increase of $|\langle a_1^\dagger a_2\rangle|$. For instance, $|\langle a_1^\dagger a_2\rangle|\approx0.453$ at $U/J=20$, which is close to the hard-core limit in Fig.~3(a) of the main text.

When the nearest-neighbor interaction $V\neq0$, the decay constant $\xi_z$ of the CDW ($n_{\bm{r}}^z\sim e^{-r/\xi_z}$) is nonzero~\cite{PhysRevB.82.134510}. As a consequence, the two pinning sites will be anti-correlated with their neighboring sites and the ground state $\ket{G}$ will be in the form $\frac{1}{\sqrt 2}(\ket{10}_{12}\ket{\eta_1}+e^{i\phi}\ket{01}_{12}\ket{\eta_2})$. Here, $\ket{\eta_1}$ and $\ket{\eta_2}$ are the states of other sites. They are approximately identical (denote as $\ket{\eta}$) when $V=0$, so that $\ket{G}\sim\frac{1}{\sqrt 2}(\ket{10}_{12}+e^{i\phi}\ket{01}_{12})\ket{\eta}$, same as Eq.~(2) of the main text. For nonzero $V$, $\ket{\eta_1}\neq\ket{\eta_2}$. The four neighboring sites of the site $1$ have a large probability to be in $\ket{0000}$ for $\ket{\eta_1}$ and $\ket{1111}$ for $\ket{\eta_2}$ (for the site $2$, exchange $\ket{0000}$ and $\ket{1111}$). These results indicate that the coherence between the two sites extends to their neighbors, rendering the entanglement of vortices in $\ket{G}$ difficult to detect.

\bibliography{References}

%merlin.mbs apsrev4-1.bst 2010-07-25 4.21a (PWD, AO, DPC) hacked
%Control: key (0)
%Control: author (72) initials jnrlst
%Control: editor formatted (1) identically to author
%Control: production of article title (-1) disabled
%Control: page (0) single
%Control: year (1) truncated
%Control: production of eprint (0) enabled
\begin{thebibliography}{85}%
\makeatletter
\providecommand \@ifxundefined [1]{%
 \@ifx{#1\undefined}
}%
\providecommand \@ifnum [1]{%
 \ifnum #1\expandafter \@firstoftwo
 \else \expandafter \@secondoftwo
 \fi
}%
\providecommand \@ifx [1]{%
 \ifx #1\expandafter \@firstoftwo
 \else \expandafter \@secondoftwo
 \fi
}%
\providecommand \natexlab [1]{#1}%
\providecommand \enquote  [1]{``#1''}%
\providecommand \bibnamefont  [1]{#1}%
\providecommand \bibfnamefont [1]{#1}%
\providecommand \citenamefont [1]{#1}%
\providecommand \href@noop [0]{\@secondoftwo}%
\providecommand \href [0]{\begingroup \@sanitize@url \@href}%
\providecommand \@href[1]{\@@startlink{#1}\@@href}%
\providecommand \@@href[1]{\endgroup#1\@@endlink}%
\providecommand \@sanitize@url [0]{\catcode `\\12\catcode `\$12\catcode
  `\&12\catcode `\#12\catcode `\^12\catcode `\_12\catcode `\%12\relax}%
\providecommand \@@startlink[1]{}%
\providecommand \@@endlink[0]{}%
\providecommand \url  [0]{\begingroup\@sanitize@url \@url }%
\providecommand \@url [1]{\endgroup\@href {#1}{\urlprefix }}%
\providecommand \urlprefix  [0]{URL }%
\providecommand \Eprint [0]{\href }%
\providecommand \doibase [0]{http://dx.doi.org/}%
\providecommand \selectlanguage [0]{\@gobble}%
\providecommand \bibinfo  [0]{\@secondoftwo}%
\providecommand \bibfield  [0]{\@secondoftwo}%
\providecommand \translation [1]{[#1]}%
\providecommand \BibitemOpen [0]{}%
\providecommand \bibitemStop [0]{}%
\providecommand \bibitemNoStop [0]{.\EOS\space}%
\providecommand \EOS [0]{\spacefactor3000\relax}%
\providecommand \BibitemShut  [1]{\csname bibitem#1\endcsname}%
\let\auto@bib@innerbib\@empty
%</preamble>
\bibitem [{\citenamefont {Wu}\ \emph {et~al.}(2004)\citenamefont {Wu},
  \citenamefont {Chen}, \citenamefont {Hu},\ and\ \citenamefont
  {Zhang}}]{PhysRevA.69.043609}%
  \BibitemOpen
  \bibfield  {author} {\bibinfo {author} {\bibfnamefont {C.}~\bibnamefont
  {Wu}}, \bibinfo {author} {\bibfnamefont {H.-d.}\ \bibnamefont {Chen}},
  \bibinfo {author} {\bibfnamefont {J.-p.}\ \bibnamefont {Hu}}, \ and\ \bibinfo
  {author} {\bibfnamefont {S.-C.}\ \bibnamefont {Zhang}},\ }\href {\doibase
  10.1103/PhysRevA.69.043609} {\bibfield  {journal} {\bibinfo  {journal} {Phys.
  Rev. A}\ }\textbf {\bibinfo {volume} {69}},\ \bibinfo {pages} {043609}
  (\bibinfo {year} {2004})}\BibitemShut {NoStop}%
\bibitem [{\citenamefont {Goldbaum}\ and\ \citenamefont
  {Mueller}(2009)}]{PhysRevA.79.021602}%
  \BibitemOpen
  \bibfield  {author} {\bibinfo {author} {\bibfnamefont {D.~S.}\ \bibnamefont
  {Goldbaum}}\ and\ \bibinfo {author} {\bibfnamefont {E.~J.}\ \bibnamefont
  {Mueller}},\ }\href {\doibase 10.1103/PhysRevA.79.021602} {\bibfield
  {journal} {\bibinfo  {journal} {Phys. Rev. A}\ }\textbf {\bibinfo {volume}
  {79}},\ \bibinfo {pages} {021602} (\bibinfo {year} {2009})}\BibitemShut
  {NoStop}%
\bibitem [{\citenamefont {Wang}\ \emph {et~al.}(2018)\citenamefont {Wang},
  \citenamefont {Edkins}, \citenamefont {Hamidian}, \citenamefont {Davis},
  \citenamefont {Fradkin},\ and\ \citenamefont
  {Kivelson}}]{PhysRevB.97.174510}%
  \BibitemOpen
  \bibfield  {author} {\bibinfo {author} {\bibfnamefont {Y.}~\bibnamefont
  {Wang}}, \bibinfo {author} {\bibfnamefont {S.~D.}\ \bibnamefont {Edkins}},
  \bibinfo {author} {\bibfnamefont {M.~H.}\ \bibnamefont {Hamidian}}, \bibinfo
  {author} {\bibfnamefont {J.~C.~S.}\ \bibnamefont {Davis}}, \bibinfo {author}
  {\bibfnamefont {E.}~\bibnamefont {Fradkin}}, \ and\ \bibinfo {author}
  {\bibfnamefont {S.~A.}\ \bibnamefont {Kivelson}},\ }\href {\doibase
  10.1103/PhysRevB.97.174510} {\bibfield  {journal} {\bibinfo  {journal} {Phys.
  Rev. B}\ }\textbf {\bibinfo {volume} {97}},\ \bibinfo {pages} {174510}
  (\bibinfo {year} {2018})}\BibitemShut {NoStop}%
\bibitem [{\citenamefont {Dai}\ \emph {et~al.}(2018)\citenamefont {Dai},
  \citenamefont {Zhang}, \citenamefont {Senthil},\ and\ \citenamefont
  {Lee}}]{PhysRevB.97.174511}%
  \BibitemOpen
  \bibfield  {author} {\bibinfo {author} {\bibfnamefont {Z.}~\bibnamefont
  {Dai}}, \bibinfo {author} {\bibfnamefont {Y.-H.}\ \bibnamefont {Zhang}},
  \bibinfo {author} {\bibfnamefont {T.}~\bibnamefont {Senthil}}, \ and\
  \bibinfo {author} {\bibfnamefont {P.~A.}\ \bibnamefont {Lee}},\ }\href
  {\doibase 10.1103/PhysRevB.97.174511} {\bibfield  {journal} {\bibinfo
  {journal} {Phys. Rev. B}\ }\textbf {\bibinfo {volume} {97}},\ \bibinfo
  {pages} {174511} (\bibinfo {year} {2018})}\BibitemShut {NoStop}%
\bibitem [{\citenamefont {Lindner}\ \emph {et~al.}(2010)\citenamefont
  {Lindner}, \citenamefont {Auerbach},\ and\ \citenamefont
  {Arovas}}]{PhysRevB.82.134510}%
  \BibitemOpen
  \bibfield  {author} {\bibinfo {author} {\bibfnamefont {N.}~\bibnamefont
  {Lindner}}, \bibinfo {author} {\bibfnamefont {A.}~\bibnamefont {Auerbach}}, \
  and\ \bibinfo {author} {\bibfnamefont {D.~P.}\ \bibnamefont {Arovas}},\
  }\href {\doibase 10.1103/PhysRevB.82.134510} {\bibfield  {journal} {\bibinfo
  {journal} {Phys. Rev. B}\ }\textbf {\bibinfo {volume} {82}},\ \bibinfo
  {pages} {134510} (\bibinfo {year} {2010})}\BibitemShut {NoStop}%
\bibitem [{\citenamefont {Ginzburg}\ and\ \citenamefont
  {Pitaevskii}(1958)}]{Ginzburg_Pitaevskii}%
  \BibitemOpen
  \bibfield  {author} {\bibinfo {author} {\bibfnamefont {V.}~\bibnamefont
  {Ginzburg}}\ and\ \bibinfo {author} {\bibfnamefont {L.}~\bibnamefont
  {Pitaevskii}},\ }\href@noop {} {\bibfield  {journal} {\bibinfo  {journal}
  {Sov. Phys. JETP}\ }\textbf {\bibinfo {volume} {7}},\ \bibinfo {pages} {858}
  (\bibinfo {year} {1958})}\BibitemShut {NoStop}%
\bibitem [{\citenamefont {Gross}(1961)}]{Gross1961}%
  \BibitemOpen
  \bibfield  {author} {\bibinfo {author} {\bibfnamefont {E.~P.}\ \bibnamefont
  {Gross}},\ }\href {\doibase 10.1007/BF02731494} {\bibfield  {journal}
  {\bibinfo  {journal} {Il Nuovo Cimento}\ }\textbf {\bibinfo {volume} {20}},\
  \bibinfo {pages} {454} (\bibinfo {year} {1961})}\BibitemShut {NoStop}%
\bibitem [{\citenamefont {Pitaevskii}(1961)}]{Pitaevskii1961}%
  \BibitemOpen
  \bibfield  {author} {\bibinfo {author} {\bibfnamefont {L.}~\bibnamefont
  {Pitaevskii}},\ }\href@noop {} {\bibfield  {journal} {\bibinfo  {journal}
  {Sov. Phys. JETP}\ }\textbf {\bibinfo {volume} {13}},\ \bibinfo {pages} {451}
  (\bibinfo {year} {1961})}\BibitemShut {NoStop}%
\bibitem [{\citenamefont {Fisher}\ and\ \citenamefont
  {Lee}(1989)}]{PhysRevB.39.2756}%
  \BibitemOpen
  \bibfield  {author} {\bibinfo {author} {\bibfnamefont {M.~P.~A.}\
  \bibnamefont {Fisher}}\ and\ \bibinfo {author} {\bibfnamefont {D.~H.}\
  \bibnamefont {Lee}},\ }\href {\doibase 10.1103/PhysRevB.39.2756} {\bibfield
  {journal} {\bibinfo  {journal} {Phys. Rev. B}\ }\textbf {\bibinfo {volume}
  {39}},\ \bibinfo {pages} {2756} (\bibinfo {year} {1989})}\BibitemShut
  {NoStop}%
\bibitem [{\citenamefont {Lee}\ and\ \citenamefont
  {Fisher}(1989)}]{PhysRevLett.63.903}%
  \BibitemOpen
  \bibfield  {author} {\bibinfo {author} {\bibfnamefont {D.-H.}\ \bibnamefont
  {Lee}}\ and\ \bibinfo {author} {\bibfnamefont {M.~P.~A.}\ \bibnamefont
  {Fisher}},\ }\href {\doibase 10.1103/PhysRevLett.63.903} {\bibfield
  {journal} {\bibinfo  {journal} {Phys. Rev. Lett.}\ }\textbf {\bibinfo
  {volume} {63}},\ \bibinfo {pages} {903} (\bibinfo {year} {1989})}\BibitemShut
  {NoStop}%
\bibitem [{\citenamefont {Bartosch}\ \emph {et~al.}(2006)\citenamefont
  {Bartosch}, \citenamefont {Balents},\ and\ \citenamefont
  {Sachdev}}]{duality2}%
  \BibitemOpen
  \bibfield  {author} {\bibinfo {author} {\bibfnamefont {L.}~\bibnamefont
  {Bartosch}}, \bibinfo {author} {\bibfnamefont {L.}~\bibnamefont {Balents}}, \
  and\ \bibinfo {author} {\bibfnamefont {S.}~\bibnamefont {Sachdev}},\ }\href
  {\doibase 10.1016/j.aop.2006.04.001} {\bibfield  {journal} {\bibinfo
  {journal} {Ann. Phys. (N.Y.)}\ }\textbf {\bibinfo {volume} {321}},\ \bibinfo
  {pages} {1528 } (\bibinfo {year} {2006})}\BibitemShut {NoStop}%
\bibitem [{\citenamefont {Gazit}\ \emph {et~al.}(2014)\citenamefont {Gazit},
  \citenamefont {Podolsky},\ and\ \citenamefont
  {Auerbach}}]{PhysRevLett.113.240601}%
  \BibitemOpen
  \bibfield  {author} {\bibinfo {author} {\bibfnamefont {S.}~\bibnamefont
  {Gazit}}, \bibinfo {author} {\bibfnamefont {D.}~\bibnamefont {Podolsky}}, \
  and\ \bibinfo {author} {\bibfnamefont {A.}~\bibnamefont {Auerbach}},\ }\href
  {\doibase 10.1103/PhysRevLett.113.240601} {\bibfield  {journal} {\bibinfo
  {journal} {Phys. Rev. Lett.}\ }\textbf {\bibinfo {volume} {113}},\ \bibinfo
  {pages} {240601} (\bibinfo {year} {2014})}\BibitemShut {NoStop}%
\bibitem [{\citenamefont {Lindner}\ \emph {et~al.}(2009)\citenamefont
  {Lindner}, \citenamefont {Auerbach},\ and\ \citenamefont
  {Arovas}}]{PhysRevLett.102.070403}%
  \BibitemOpen
  \bibfield  {author} {\bibinfo {author} {\bibfnamefont {N.~H.}\ \bibnamefont
  {Lindner}}, \bibinfo {author} {\bibfnamefont {A.}~\bibnamefont {Auerbach}}, \
  and\ \bibinfo {author} {\bibfnamefont {D.~P.}\ \bibnamefont {Arovas}},\
  }\href {\doibase 10.1103/PhysRevLett.102.070403} {\bibfield  {journal}
  {\bibinfo  {journal} {Phys. Rev. Lett.}\ }\textbf {\bibinfo {volume} {102}},\
  \bibinfo {pages} {070403} (\bibinfo {year} {2009})}\BibitemShut {NoStop}%
\bibitem [{\citenamefont {Huber}\ and\ \citenamefont {Lindner}(2011)}]{pnas}%
  \BibitemOpen
  \bibfield  {author} {\bibinfo {author} {\bibfnamefont {S.~D.}\ \bibnamefont
  {Huber}}\ and\ \bibinfo {author} {\bibfnamefont {N.~H.}\ \bibnamefont
  {Lindner}},\ }\href {\doibase 10.1073/pnas.1110813108} {\bibfield  {journal}
  {\bibinfo  {journal} {Proc. Natl. Acad. Sci. U.S.A.}\ }\textbf {\bibinfo
  {volume} {108}},\ \bibinfo {pages} {19925} (\bibinfo {year}
  {2011})}\BibitemShut {NoStop}%
\bibitem [{\citenamefont {{Lindner}}\ \emph {et~al.}()\citenamefont
  {{Lindner}}, \citenamefont {{Auerbach}},\ and\ \citenamefont
  {{Arovas}}}]{arxiv0701571v1}%
  \BibitemOpen
  \bibfield  {author} {\bibinfo {author} {\bibfnamefont {N.~H.}\ \bibnamefont
  {{Lindner}}}, \bibinfo {author} {\bibfnamefont {A.}~\bibnamefont
  {{Auerbach}}}, \ and\ \bibinfo {author} {\bibfnamefont {D.~P.}\ \bibnamefont
  {{Arovas}}},\ }\href@noop {} {\bibfield  {journal} {\bibinfo  {journal}
  {arXiv:}\ }}\Eprint {http://arxiv.org/abs/cond-mat/0701571v1}
  {cond-mat/0701571v1} \BibitemShut {NoStop}%
\bibitem [{\citenamefont {Raman}(1928)}]{Raman1928}%
  \BibitemOpen
  \bibfield  {author} {\bibinfo {author} {\bibfnamefont {C.~V.}\ \bibnamefont
  {Raman}},\ }\href@noop {} {\bibfield  {journal} {\bibinfo  {journal} {Indian
  J. Phys.}\ }\textbf {\bibinfo {volume} {2}},\ \bibinfo {pages} {387}
  (\bibinfo {year} {1928})}\BibitemShut {NoStop}%
\bibitem [{\citenamefont {Bakr}\ \emph {et~al.}(2009)\citenamefont {Bakr},
  \citenamefont {Gillen}, \citenamefont {Peng}, \citenamefont {F\"olling},\
  and\ \citenamefont {Greiner}}]{Greiner_Nature}%
  \BibitemOpen
  \bibfield  {author} {\bibinfo {author} {\bibfnamefont {W.~S.}\ \bibnamefont
  {Bakr}}, \bibinfo {author} {\bibfnamefont {J.~I.}\ \bibnamefont {Gillen}},
  \bibinfo {author} {\bibfnamefont {A.}~\bibnamefont {Peng}}, \bibinfo {author}
  {\bibfnamefont {S.}~\bibnamefont {F\"olling}}, \ and\ \bibinfo {author}
  {\bibfnamefont {M.}~\bibnamefont {Greiner}},\ }\href
  {http://dx.doi.org/10.1038/nature08482} {\bibfield  {journal} {\bibinfo
  {journal} {Nature}\ }\textbf {\bibinfo {volume} {462}},\ \bibinfo {pages}
  {74} (\bibinfo {year} {2009})}\BibitemShut {NoStop}%
\bibitem [{\citenamefont {Bakr}\ \emph {et~al.}(2010)\citenamefont {Bakr},
  \citenamefont {Peng}, \citenamefont {Tai}, \citenamefont {Ma}, \citenamefont
  {Simon}, \citenamefont {Gillen}, \citenamefont {F{\"o}lling}, \citenamefont
  {Pollet},\ and\ \citenamefont {Greiner}}]{Bakr547}%
  \BibitemOpen
  \bibfield  {author} {\bibinfo {author} {\bibfnamefont {W.~S.}\ \bibnamefont
  {Bakr}}, \bibinfo {author} {\bibfnamefont {A.}~\bibnamefont {Peng}}, \bibinfo
  {author} {\bibfnamefont {M.~E.}\ \bibnamefont {Tai}}, \bibinfo {author}
  {\bibfnamefont {R.}~\bibnamefont {Ma}}, \bibinfo {author} {\bibfnamefont
  {J.}~\bibnamefont {Simon}}, \bibinfo {author} {\bibfnamefont {J.~I.}\
  \bibnamefont {Gillen}}, \bibinfo {author} {\bibfnamefont {S.}~\bibnamefont
  {F{\"o}lling}}, \bibinfo {author} {\bibfnamefont {L.}~\bibnamefont {Pollet}},
  \ and\ \bibinfo {author} {\bibfnamefont {M.}~\bibnamefont {Greiner}},\ }\href
  {\doibase 10.1126/science.1192368} {\bibfield  {journal} {\bibinfo  {journal}
  {Science}\ }\textbf {\bibinfo {volume} {329}},\ \bibinfo {pages} {547}
  (\bibinfo {year} {2010})}\BibitemShut {NoStop}%
\bibitem [{\citenamefont {Hwa}\ \emph {et~al.}(1993)\citenamefont {Hwa},
  \citenamefont {Le~Doussal}, \citenamefont {Nelson},\ and\ \citenamefont
  {Vinokur}}]{PhysRevLett.71.3545}%
  \BibitemOpen
  \bibfield  {author} {\bibinfo {author} {\bibfnamefont {T.}~\bibnamefont
  {Hwa}}, \bibinfo {author} {\bibfnamefont {P.}~\bibnamefont {Le~Doussal}},
  \bibinfo {author} {\bibfnamefont {D.~R.}\ \bibnamefont {Nelson}}, \ and\
  \bibinfo {author} {\bibfnamefont {V.~M.}\ \bibnamefont {Vinokur}},\ }\href
  {\doibase 10.1103/PhysRevLett.71.3545} {\bibfield  {journal} {\bibinfo
  {journal} {Phys. Rev. Lett.}\ }\textbf {\bibinfo {volume} {71}},\ \bibinfo
  {pages} {3545} (\bibinfo {year} {1993})}\BibitemShut {NoStop}%
\bibitem [{\citenamefont {Kato}\ \emph {et~al.}(2008)\citenamefont {Kato},
  \citenamefont {Shibauchi}, \citenamefont {Matsuda}, \citenamefont
  {Thompson},\ and\ \citenamefont {Krusin-Elbaum}}]{PhysRevLett.101.027003}%
  \BibitemOpen
  \bibfield  {author} {\bibinfo {author} {\bibfnamefont {T.}~\bibnamefont
  {Kato}}, \bibinfo {author} {\bibfnamefont {T.}~\bibnamefont {Shibauchi}},
  \bibinfo {author} {\bibfnamefont {Y.}~\bibnamefont {Matsuda}}, \bibinfo
  {author} {\bibfnamefont {J.~R.}\ \bibnamefont {Thompson}}, \ and\ \bibinfo
  {author} {\bibfnamefont {L.}~\bibnamefont {Krusin-Elbaum}},\ }\href {\doibase
  10.1103/PhysRevLett.101.027003} {\bibfield  {journal} {\bibinfo  {journal}
  {Phys. Rev. Lett.}\ }\textbf {\bibinfo {volume} {101}},\ \bibinfo {pages}
  {027003} (\bibinfo {year} {2008})}\BibitemShut {NoStop}%
\bibitem [{\citenamefont {Nelson}(1988)}]{PhysRevLett.60.1973}%
  \BibitemOpen
  \bibfield  {author} {\bibinfo {author} {\bibfnamefont {D.~R.}\ \bibnamefont
  {Nelson}},\ }\href {\doibase 10.1103/PhysRevLett.60.1973} {\bibfield
  {journal} {\bibinfo  {journal} {Phys. Rev. Lett.}\ }\textbf {\bibinfo
  {volume} {60}},\ \bibinfo {pages} {1973} (\bibinfo {year}
  {1988})}\BibitemShut {NoStop}%
\bibitem [{\citenamefont {Sch\"onenberger}\ \emph {et~al.}(1995)\citenamefont
  {Sch\"onenberger}, \citenamefont {Geshkenbein},\ and\ \citenamefont
  {Blatter}}]{PhysRevLett.75.1380}%
  \BibitemOpen
  \bibfield  {author} {\bibinfo {author} {\bibfnamefont {A.}~\bibnamefont
  {Sch\"onenberger}}, \bibinfo {author} {\bibfnamefont {V.}~\bibnamefont
  {Geshkenbein}}, \ and\ \bibinfo {author} {\bibfnamefont {G.}~\bibnamefont
  {Blatter}},\ }\href {\doibase 10.1103/PhysRevLett.75.1380} {\bibfield
  {journal} {\bibinfo  {journal} {Phys. Rev. Lett.}\ }\textbf {\bibinfo
  {volume} {75}},\ \bibinfo {pages} {1380} (\bibinfo {year}
  {1995})}\BibitemShut {NoStop}%
\bibitem [{\citenamefont {Chevalier}(1995)}]{Neutron-Stars}%
  \BibitemOpen
  \bibfield  {author} {\bibinfo {author} {\bibfnamefont {E.}~\bibnamefont
  {Chevalier}},\ }\href {http://stacks.iop.org/0295-5075/29/i=2/a=013}
  {\bibfield  {journal} {\bibinfo  {journal} {EPL (Europhysics Letters)}\
  }\textbf {\bibinfo {volume} {29}},\ \bibinfo {pages} {181} (\bibinfo {year}
  {1995})}\BibitemShut {NoStop}%
\bibitem [{\citenamefont {Cooper}\ \emph {et~al.}(2001)\citenamefont {Cooper},
  \citenamefont {Wilkin},\ and\ \citenamefont {Gunn}}]{PhysRevLett.87.120405}%
  \BibitemOpen
  \bibfield  {author} {\bibinfo {author} {\bibfnamefont {N.~R.}\ \bibnamefont
  {Cooper}}, \bibinfo {author} {\bibfnamefont {N.~K.}\ \bibnamefont {Wilkin}},
  \ and\ \bibinfo {author} {\bibfnamefont {J.~M.~F.}\ \bibnamefont {Gunn}},\
  }\href {\doibase 10.1103/PhysRevLett.87.120405} {\bibfield  {journal}
  {\bibinfo  {journal} {Phys. Rev. Lett.}\ }\textbf {\bibinfo {volume} {87}},\
  \bibinfo {pages} {120405} (\bibinfo {year} {2001})}\BibitemShut {NoStop}%
\bibitem [{\citenamefont {Liu}\ \emph {et~al.}(2011)\citenamefont {Liu},
  \citenamefont {Guo}, \citenamefont {Vedral},\ and\ \citenamefont
  {Fan}}]{PhysRevA.83.013620}%
  \BibitemOpen
  \bibfield  {author} {\bibinfo {author} {\bibfnamefont {Z.}~\bibnamefont
  {Liu}}, \bibinfo {author} {\bibfnamefont {H.-L.}\ \bibnamefont {Guo}},
  \bibinfo {author} {\bibfnamefont {V.}~\bibnamefont {Vedral}}, \ and\ \bibinfo
  {author} {\bibfnamefont {H.}~\bibnamefont {Fan}},\ }\href {\doibase
  10.1103/PhysRevA.83.013620} {\bibfield  {journal} {\bibinfo  {journal} {Phys.
  Rev. A}\ }\textbf {\bibinfo {volume} {83}},\ \bibinfo {pages} {013620}
  (\bibinfo {year} {2011})}\BibitemShut {NoStop}%
\bibitem [{\citenamefont {Lo~Gullo}\ \emph {et~al.}(2010)\citenamefont
  {Lo~Gullo}, \citenamefont {McEndoo}, \citenamefont {Busch},\ and\
  \citenamefont {Paternostro}}]{PhysRevA.81.053625}%
  \BibitemOpen
  \bibfield  {author} {\bibinfo {author} {\bibfnamefont {N.}~\bibnamefont
  {Lo~Gullo}}, \bibinfo {author} {\bibfnamefont {S.}~\bibnamefont {McEndoo}},
  \bibinfo {author} {\bibfnamefont {T.}~\bibnamefont {Busch}}, \ and\ \bibinfo
  {author} {\bibfnamefont {M.}~\bibnamefont {Paternostro}},\ }\href {\doibase
  10.1103/PhysRevA.81.053625} {\bibfield  {journal} {\bibinfo  {journal} {Phys.
  Rev. A}\ }\textbf {\bibinfo {volume} {81}},\ \bibinfo {pages} {053625}
  (\bibinfo {year} {2010})}\BibitemShut {NoStop}%
\bibitem [{\citenamefont {Vedral}(2008)}]{Vedral_macro}%
  \BibitemOpen
  \bibfield  {author} {\bibinfo {author} {\bibfnamefont {V.}~\bibnamefont
  {Vedral}},\ }\href {http://dx.doi.org/10.1038/nature07124} {\bibfield
  {journal} {\bibinfo  {journal} {Nature}\ }\textbf {\bibinfo {volume} {453}},\
  \bibinfo {pages} {1004} (\bibinfo {year} {2008})}\BibitemShut {NoStop}%
\bibitem [{\citenamefont {Alicki}\ \emph {et~al.}(2008)\citenamefont {Alicki},
  \citenamefont {Piani},\ and\ \citenamefont {Ryn}}]{Quantumness1}%
  \BibitemOpen
  \bibfield  {author} {\bibinfo {author} {\bibfnamefont {R.}~\bibnamefont
  {Alicki}}, \bibinfo {author} {\bibfnamefont {M.}~\bibnamefont {Piani}}, \
  and\ \bibinfo {author} {\bibfnamefont {N.~V.}\ \bibnamefont {Ryn}},\ }\href
  {https://doi.org/10.1088/1751-8113/41/49/495303} {\bibfield  {journal}
  {\bibinfo  {journal} {J. Phys. A: Math. Theor.}\ }\textbf {\bibinfo {volume}
  {41}},\ \bibinfo {pages} {495303} (\bibinfo {year} {2008})}\BibitemShut
  {NoStop}%
\bibitem [{\citenamefont {Modi}\ \emph {et~al.}(2012)\citenamefont {Modi},
  \citenamefont {Fazio}, \citenamefont {Pascazio}, \citenamefont {Vedral},\
  and\ \citenamefont {Yuasa}}]{Quantumness2}%
  \BibitemOpen
  \bibfield  {author} {\bibinfo {author} {\bibfnamefont {K.}~\bibnamefont
  {Modi}}, \bibinfo {author} {\bibfnamefont {R.}~\bibnamefont {Fazio}},
  \bibinfo {author} {\bibfnamefont {S.}~\bibnamefont {Pascazio}}, \bibinfo
  {author} {\bibfnamefont {V.}~\bibnamefont {Vedral}}, \ and\ \bibinfo {author}
  {\bibfnamefont {K.}~\bibnamefont {Yuasa}},\ }\href {\doibase
  10.1098/rsta.2011.0353} {\bibfield  {journal} {\bibinfo  {journal} {Phil.
  Trans. R. Soc. A}\ }\textbf {\bibinfo {volume} {370}},\ \bibinfo {pages}
  {4810} (\bibinfo {year} {2012})}\BibitemShut {NoStop}%
\bibitem [{\citenamefont {Hillery}\ \emph {et~al.}(1999)\citenamefont
  {Hillery}, \citenamefont {Bu\ifmmode~\check{z}\else \v{z}\fi{}ek},\ and\
  \citenamefont {Berthiaume}}]{PhysRevA.59.1829}%
  \BibitemOpen
  \bibfield  {author} {\bibinfo {author} {\bibfnamefont {M.}~\bibnamefont
  {Hillery}}, \bibinfo {author} {\bibfnamefont {V.}~\bibnamefont
  {Bu\ifmmode~\check{z}\else \v{z}\fi{}ek}}, \ and\ \bibinfo {author}
  {\bibfnamefont {A.}~\bibnamefont {Berthiaume}},\ }\href {\doibase
  10.1103/PhysRevA.59.1829} {\bibfield  {journal} {\bibinfo  {journal} {Phys.
  Rev. A}\ }\textbf {\bibinfo {volume} {59}},\ \bibinfo {pages} {1829}
  (\bibinfo {year} {1999})}\BibitemShut {NoStop}%
\bibitem [{\citenamefont {Gilchrist}\ \emph {et~al.}(2004)\citenamefont
  {Gilchrist}, \citenamefont {Nemoto}, \citenamefont {Munro}, \citenamefont
  {Ralph}, \citenamefont {Glancy}, \citenamefont {Braunstein},\ and\
  \citenamefont {Milburn}}]{cats2}%
  \BibitemOpen
  \bibfield  {author} {\bibinfo {author} {\bibfnamefont {A.}~\bibnamefont
  {Gilchrist}}, \bibinfo {author} {\bibfnamefont {K.}~\bibnamefont {Nemoto}},
  \bibinfo {author} {\bibfnamefont {W.~J.}\ \bibnamefont {Munro}}, \bibinfo
  {author} {\bibfnamefont {T.~C.}\ \bibnamefont {Ralph}}, \bibinfo {author}
  {\bibfnamefont {S.}~\bibnamefont {Glancy}}, \bibinfo {author} {\bibfnamefont
  {S.~L.}\ \bibnamefont {Braunstein}}, \ and\ \bibinfo {author} {\bibfnamefont
  {G.~J.}\ \bibnamefont {Milburn}},\ }\href
  {http://dx.doi.org/10.1088/1464-4266/6/8/032} {\bibfield  {journal} {\bibinfo
   {journal} {J. Opt. B}\ }\textbf {\bibinfo {volume} {6}},\ \bibinfo {pages}
  {S828} (\bibinfo {year} {2004})}\BibitemShut {NoStop}%
\bibitem [{\citenamefont {Lee}\ \emph {et~al.}(2002)\citenamefont {Lee},
  \citenamefont {Kok},\ and\ \citenamefont {Dowling}}]{cats3}%
  \BibitemOpen
  \bibfield  {author} {\bibinfo {author} {\bibfnamefont {H.}~\bibnamefont
  {Lee}}, \bibinfo {author} {\bibfnamefont {P.}~\bibnamefont {Kok}}, \ and\
  \bibinfo {author} {\bibfnamefont {J.~P.}\ \bibnamefont {Dowling}},\ }\href
  {\doibase 10.1080/0950034021000011536} {\bibfield  {journal} {\bibinfo
  {journal} {J. Mod. Opt.}\ }\textbf {\bibinfo {volume} {49}},\ \bibinfo
  {pages} {2325} (\bibinfo {year} {2002})}\BibitemShut {NoStop}%
\bibitem [{\citenamefont {Schr{\"o}dinger}(1935)}]{Schrödinger1935}%
  \BibitemOpen
  \bibfield  {author} {\bibinfo {author} {\bibfnamefont {E.}~\bibnamefont
  {Schr{\"o}dinger}},\ }\href {\doibase 10.1007/BF01491891} {\bibfield
  {journal} {\bibinfo  {journal} {Naturwissenschaften}\ }\textbf {\bibinfo
  {volume} {23}},\ \bibinfo {pages} {807} (\bibinfo {year} {1935})}\BibitemShut
  {NoStop}%
\bibitem [{\citenamefont {Cirac}\ \emph {et~al.}(1998)\citenamefont {Cirac},
  \citenamefont {Lewenstein}, \citenamefont {M\o{}lmer},\ and\ \citenamefont
  {Zoller}}]{PhysRevA.57.1208}%
  \BibitemOpen
  \bibfield  {author} {\bibinfo {author} {\bibfnamefont {J.~I.}\ \bibnamefont
  {Cirac}}, \bibinfo {author} {\bibfnamefont {M.}~\bibnamefont {Lewenstein}},
  \bibinfo {author} {\bibfnamefont {K.}~\bibnamefont {M\o{}lmer}}, \ and\
  \bibinfo {author} {\bibfnamefont {P.}~\bibnamefont {Zoller}},\ }\href
  {\doibase 10.1103/PhysRevA.57.1208} {\bibfield  {journal} {\bibinfo
  {journal} {Phys. Rev. A}\ }\textbf {\bibinfo {volume} {57}},\ \bibinfo
  {pages} {1208} (\bibinfo {year} {1998})}\BibitemShut {NoStop}%
\bibitem [{\citenamefont {Ho}\ and\ \citenamefont {Ciobanu}(2004)}]{Ho2004}%
  \BibitemOpen
  \bibfield  {author} {\bibinfo {author} {\bibfnamefont {T.-L.}\ \bibnamefont
  {Ho}}\ and\ \bibinfo {author} {\bibfnamefont {C.~V.}\ \bibnamefont
  {Ciobanu}},\ }\href {\doibase 10.1023/B:JOLT.0000024552.87247.eb} {\bibfield
  {journal} {\bibinfo  {journal} {J. Low Temp. Phys.}\ }\textbf {\bibinfo
  {volume} {135}},\ \bibinfo {pages} {257} (\bibinfo {year}
  {2004})}\BibitemShut {NoStop}%
\bibitem [{\citenamefont {Wang}(2007)}]{PhysRevLett.98.060403}%
  \BibitemOpen
  \bibfield  {author} {\bibinfo {author} {\bibfnamefont {D.-W.}\ \bibnamefont
  {Wang}},\ }\href {\doibase 10.1103/PhysRevLett.98.060403} {\bibfield
  {journal} {\bibinfo  {journal} {Phys. Rev. Lett.}\ }\textbf {\bibinfo
  {volume} {98}},\ \bibinfo {pages} {060403} (\bibinfo {year}
  {2007})}\BibitemShut {NoStop}%
\bibitem [{\citenamefont {Hald}\ \emph {et~al.}(1999)\citenamefont {Hald},
  \citenamefont {S\o{}rensen}, \citenamefont {Schori},\ and\ \citenamefont
  {Polzik}}]{PhysRevLett.83.1319}%
  \BibitemOpen
  \bibfield  {author} {\bibinfo {author} {\bibfnamefont {J.}~\bibnamefont
  {Hald}}, \bibinfo {author} {\bibfnamefont {J.~L.}\ \bibnamefont
  {S\o{}rensen}}, \bibinfo {author} {\bibfnamefont {C.}~\bibnamefont {Schori}},
  \ and\ \bibinfo {author} {\bibfnamefont {E.~S.}\ \bibnamefont {Polzik}},\
  }\href {\doibase 10.1103/PhysRevLett.83.1319} {\bibfield  {journal} {\bibinfo
   {journal} {Phys. Rev. Lett.}\ }\textbf {\bibinfo {volume} {83}},\ \bibinfo
  {pages} {1319} (\bibinfo {year} {1999})}\BibitemShut {NoStop}%
\bibitem [{\citenamefont {S\o{}rensen}\ \emph {et~al.}(2001)\citenamefont
  {S\o{}rensen}, \citenamefont {Duan}, \citenamefont {Cirac},\ and\
  \citenamefont {Zoller}}]{etg_Nature}%
  \BibitemOpen
  \bibfield  {author} {\bibinfo {author} {\bibfnamefont {A.}~\bibnamefont
  {S\o{}rensen}}, \bibinfo {author} {\bibfnamefont {L.-M.}\ \bibnamefont
  {Duan}}, \bibinfo {author} {\bibfnamefont {J.~I.}\ \bibnamefont {Cirac}}, \
  and\ \bibinfo {author} {\bibfnamefont {P.}~\bibnamefont {Zoller}},\ }\href
  {http://dx.doi.org/10.1038/35051038} {\bibfield  {journal} {\bibinfo
  {journal} {Nature}\ }\textbf {\bibinfo {volume} {409}},\ \bibinfo {pages}
  {63} (\bibinfo {year} {2001})}\BibitemShut {NoStop}%
\bibitem [{\citenamefont {Julsgaard}\ \emph {et~al.}(2001)\citenamefont
  {Julsgaard}, \citenamefont {Kozhekin},\ and\ \citenamefont
  {Polzik}}]{Julsgaard_Nature}%
  \BibitemOpen
  \bibfield  {author} {\bibinfo {author} {\bibfnamefont {B.}~\bibnamefont
  {Julsgaard}}, \bibinfo {author} {\bibfnamefont {A.}~\bibnamefont {Kozhekin}},
  \ and\ \bibinfo {author} {\bibfnamefont {E.~S.}\ \bibnamefont {Polzik}},\
  }\href {http://dx.doi.org/10.1038/35096524} {\bibfield  {journal} {\bibinfo
  {journal} {Nature}\ }\textbf {\bibinfo {volume} {413}},\ \bibinfo {pages}
  {400} (\bibinfo {year} {2001})}\BibitemShut {NoStop}%
\bibitem [{\citenamefont {Micheli}\ \emph {et~al.}(2003)\citenamefont
  {Micheli}, \citenamefont {Jaksch}, \citenamefont {Cirac},\ and\ \citenamefont
  {Zoller}}]{PhysRevA.67.013607}%
  \BibitemOpen
  \bibfield  {author} {\bibinfo {author} {\bibfnamefont {A.}~\bibnamefont
  {Micheli}}, \bibinfo {author} {\bibfnamefont {D.}~\bibnamefont {Jaksch}},
  \bibinfo {author} {\bibfnamefont {J.~I.}\ \bibnamefont {Cirac}}, \ and\
  \bibinfo {author} {\bibfnamefont {P.}~\bibnamefont {Zoller}},\ }\href
  {\doibase 10.1103/PhysRevA.67.013607} {\bibfield  {journal} {\bibinfo
  {journal} {Phys. Rev. A}\ }\textbf {\bibinfo {volume} {67}},\ \bibinfo
  {pages} {013607} (\bibinfo {year} {2003})}\BibitemShut {NoStop}%
\bibitem [{\citenamefont {You}(2003)}]{PhysRevLett.90.030402}%
  \BibitemOpen
  \bibfield  {author} {\bibinfo {author} {\bibfnamefont {L.}~\bibnamefont
  {You}},\ }\href {\doibase 10.1103/PhysRevLett.90.030402} {\bibfield
  {journal} {\bibinfo  {journal} {Phys. Rev. Lett.}\ }\textbf {\bibinfo
  {volume} {90}},\ \bibinfo {pages} {030402} (\bibinfo {year}
  {2003})}\BibitemShut {NoStop}%
\bibitem [{\citenamefont {Est\`eve}\ \emph {et~al.}(2008)\citenamefont
  {Est\`eve}, \citenamefont {Gross}, \citenamefont {Weller}, \citenamefont
  {Giovanazzi},\ and\ \citenamefont {Oberthaler}}]{Oberthaler_Nature}%
  \BibitemOpen
  \bibfield  {author} {\bibinfo {author} {\bibfnamefont {J.}~\bibnamefont
  {Est\`eve}}, \bibinfo {author} {\bibfnamefont {C.}~\bibnamefont {Gross}},
  \bibinfo {author} {\bibfnamefont {A.}~\bibnamefont {Weller}}, \bibinfo
  {author} {\bibfnamefont {S.}~\bibnamefont {Giovanazzi}}, \ and\ \bibinfo
  {author} {\bibfnamefont {M.~K.}\ \bibnamefont {Oberthaler}},\ }\href
  {http://dx.doi.org/10.1038/nature07332} {\bibfield  {journal} {\bibinfo
  {journal} {Nature}\ }\textbf {\bibinfo {volume} {455}},\ \bibinfo {pages}
  {1216} (\bibinfo {year} {2008})}\BibitemShut {NoStop}%
\bibitem [{sup()}]{supp}%
  \BibitemOpen
  \href@noop {} {}\bibinfo {note} {See Supplemental Material for the
  interpretation of the vortex pinning effect, estimation of the background
  vorticity, the entangled state of vortices, the coherent charge-density
  waves, the phase of $\langle a_1^\dagger a_2\rangle$, perturbation theory,
  entangled qubits, and finite on-site and nearest-neighbor
  interactions.}\BibitemShut {Stop}%
\bibitem [{vor()}]{vorticity_deg}%
  \BibitemOpen
  \href@noop {} {}\bibinfo {note} {The ground states for
  $J_{\textrm{pin}}/J=1,0.9$ are doubly degenerate and $\langle\nabla\times\bm
  j\rangle$ is averaged over all superpositions of the ground states. The
  statistical deviations are very small ($\approx0.015,0.021$ for
  $J_{\textrm{pin}}/J=1,0.9$ respectively).}\BibitemShut {Stop}%
\bibitem [{\citenamefont {Jaksch}\ and\ \citenamefont
  {Zoller}(2003)}]{Jaksch_Zoller}%
  \BibitemOpen
  \bibfield  {author} {\bibinfo {author} {\bibfnamefont {D.}~\bibnamefont
  {Jaksch}}\ and\ \bibinfo {author} {\bibfnamefont {P.}~\bibnamefont
  {Zoller}},\ }\href {http://stacks.iop.org/1367-2630/5/i=1/a=356} {\bibfield
  {journal} {\bibinfo  {journal} {New J. Phys.}\ }\textbf {\bibinfo {volume}
  {5}},\ \bibinfo {pages} {56} (\bibinfo {year} {2003})}\BibitemShut {NoStop}%
\bibitem [{\citenamefont {Gerbier}\ and\ \citenamefont
  {Dalibard}(2010)}]{Gerbier_Dalibard}%
  \BibitemOpen
  \bibfield  {author} {\bibinfo {author} {\bibfnamefont {F.}~\bibnamefont
  {Gerbier}}\ and\ \bibinfo {author} {\bibfnamefont {J.}~\bibnamefont
  {Dalibard}},\ }\href {http://stacks.iop.org/1367-2630/12/i=3/a=033007}
  {\bibfield  {journal} {\bibinfo  {journal} {New J. Phys.}\ }\textbf {\bibinfo
  {volume} {12}},\ \bibinfo {pages} {033007} (\bibinfo {year}
  {2010})}\BibitemShut {NoStop}%
\bibitem [{\citenamefont {S\o{}rensen}\ \emph {et~al.}(2005)\citenamefont
  {S\o{}rensen}, \citenamefont {Demler},\ and\ \citenamefont
  {Lukin}}]{PhysRevLett.94.086803}%
  \BibitemOpen
  \bibfield  {author} {\bibinfo {author} {\bibfnamefont {A.~S.}\ \bibnamefont
  {S\o{}rensen}}, \bibinfo {author} {\bibfnamefont {E.}~\bibnamefont {Demler}},
  \ and\ \bibinfo {author} {\bibfnamefont {M.~D.}\ \bibnamefont {Lukin}},\
  }\href {\doibase 10.1103/PhysRevLett.94.086803} {\bibfield  {journal}
  {\bibinfo  {journal} {Phys. Rev. Lett.}\ }\textbf {\bibinfo {volume} {94}},\
  \bibinfo {pages} {086803} (\bibinfo {year} {2005})}\BibitemShut {NoStop}%
\bibitem [{\citenamefont {Aidelsburger}\ \emph {et~al.}(2013)\citenamefont
  {Aidelsburger}, \citenamefont {Atala}, \citenamefont {Lohse}, \citenamefont
  {Barreiro}, \citenamefont {Paredes},\ and\ \citenamefont
  {Bloch}}]{PhysRevLett.111.185301}%
  \BibitemOpen
  \bibfield  {author} {\bibinfo {author} {\bibfnamefont {M.}~\bibnamefont
  {Aidelsburger}}, \bibinfo {author} {\bibfnamefont {M.}~\bibnamefont {Atala}},
  \bibinfo {author} {\bibfnamefont {M.}~\bibnamefont {Lohse}}, \bibinfo
  {author} {\bibfnamefont {J.~T.}\ \bibnamefont {Barreiro}}, \bibinfo {author}
  {\bibfnamefont {B.}~\bibnamefont {Paredes}}, \ and\ \bibinfo {author}
  {\bibfnamefont {I.}~\bibnamefont {Bloch}},\ }\href {\doibase
  10.1103/PhysRevLett.111.185301} {\bibfield  {journal} {\bibinfo  {journal}
  {Phys. Rev. Lett.}\ }\textbf {\bibinfo {volume} {111}},\ \bibinfo {pages}
  {185301} (\bibinfo {year} {2013})}\BibitemShut {NoStop}%
\bibitem [{\citenamefont {Miyake}\ \emph {et~al.}(2013)\citenamefont {Miyake},
  \citenamefont {Siviloglou}, \citenamefont {Kennedy}, \citenamefont {Burton},\
  and\ \citenamefont {Ketterle}}]{PhysRevLett.111.185302}%
  \BibitemOpen
  \bibfield  {author} {\bibinfo {author} {\bibfnamefont {H.}~\bibnamefont
  {Miyake}}, \bibinfo {author} {\bibfnamefont {G.~A.}\ \bibnamefont
  {Siviloglou}}, \bibinfo {author} {\bibfnamefont {C.~J.}\ \bibnamefont
  {Kennedy}}, \bibinfo {author} {\bibfnamefont {W.~C.}\ \bibnamefont {Burton}},
  \ and\ \bibinfo {author} {\bibfnamefont {W.}~\bibnamefont {Ketterle}},\
  }\href {\doibase 10.1103/PhysRevLett.111.185302} {\bibfield  {journal}
  {\bibinfo  {journal} {Phys. Rev. Lett.}\ }\textbf {\bibinfo {volume} {111}},\
  \bibinfo {pages} {185302} (\bibinfo {year} {2013})}\BibitemShut {NoStop}%
\bibitem [{\citenamefont {Lin}\ \emph {et~al.}(2009)\citenamefont {Lin},
  \citenamefont {Compton}, \citenamefont {Jim\'enez-Garc\'{\i}a}, \citenamefont
  {Porto},\ and\ \citenamefont {Spielman}}]{Lin_Nature}%
  \BibitemOpen
  \bibfield  {author} {\bibinfo {author} {\bibfnamefont {Y.-J.}\ \bibnamefont
  {Lin}}, \bibinfo {author} {\bibfnamefont {R.~L.}\ \bibnamefont {Compton}},
  \bibinfo {author} {\bibfnamefont {K.}~\bibnamefont {Jim\'enez-Garc\'{\i}a}},
  \bibinfo {author} {\bibfnamefont {J.~V.}\ \bibnamefont {Porto}}, \ and\
  \bibinfo {author} {\bibfnamefont {I.~B.}\ \bibnamefont {Spielman}},\ }\href
  {http://dx.doi.org/10.1038/nature08609} {\bibfield  {journal} {\bibinfo
  {journal} {Nature (London)}\ }\textbf {\bibinfo {volume} {462}},\ \bibinfo
  {pages} {628} (\bibinfo {year} {2009})}\BibitemShut {NoStop}%
\bibitem [{\citenamefont {Aidelsburger}\ \emph {et~al.}(2014)\citenamefont
  {Aidelsburger}, \citenamefont {Lohse}, \citenamefont {Schweizer},
  \citenamefont {Atala}, \citenamefont {Barreiro}, \citenamefont
  {Nascimb\`ene}, \citenamefont {Cooper}, \citenamefont {Bloch},\ and\
  \citenamefont {Goldman}}]{Aidelsburger}%
  \BibitemOpen
  \bibfield  {author} {\bibinfo {author} {\bibfnamefont {M.}~\bibnamefont
  {Aidelsburger}}, \bibinfo {author} {\bibfnamefont {M.}~\bibnamefont {Lohse}},
  \bibinfo {author} {\bibfnamefont {C.}~\bibnamefont {Schweizer}}, \bibinfo
  {author} {\bibfnamefont {M.}~\bibnamefont {Atala}}, \bibinfo {author}
  {\bibfnamefont {J.~T.}\ \bibnamefont {Barreiro}}, \bibinfo {author}
  {\bibfnamefont {S.}~\bibnamefont {Nascimb\`ene}}, \bibinfo {author}
  {\bibfnamefont {N.~R.}\ \bibnamefont {Cooper}}, \bibinfo {author}
  {\bibfnamefont {I.}~\bibnamefont {Bloch}}, \ and\ \bibinfo {author}
  {\bibfnamefont {N.}~\bibnamefont {Goldman}},\ }\href
  {http://dx.doi.org/10.1038/nphys3171} {\bibfield  {journal} {\bibinfo
  {journal} {Nature Phys.}\ }\textbf {\bibinfo {volume} {11}},\ \bibinfo
  {pages} {162} (\bibinfo {year} {2014})}\BibitemShut {NoStop}%
\bibitem [{\citenamefont {Altman}\ and\ \citenamefont
  {Auerbach}(1998)}]{PhysRevLett.81.4484}%
  \BibitemOpen
  \bibfield  {author} {\bibinfo {author} {\bibfnamefont {E.}~\bibnamefont
  {Altman}}\ and\ \bibinfo {author} {\bibfnamefont {A.}~\bibnamefont
  {Auerbach}},\ }\href {\doibase 10.1103/PhysRevLett.81.4484} {\bibfield
  {journal} {\bibinfo  {journal} {Phys. Rev. Lett.}\ }\textbf {\bibinfo
  {volume} {81}},\ \bibinfo {pages} {4484} (\bibinfo {year}
  {1998})}\BibitemShut {NoStop}%
\bibitem [{\citenamefont {Fazio}\ and\ \citenamefont {van~der
  Zant}(2001)}]{FAZIO2001235}%
  \BibitemOpen
  \bibfield  {author} {\bibinfo {author} {\bibfnamefont {R.}~\bibnamefont
  {Fazio}}\ and\ \bibinfo {author} {\bibfnamefont {H.}~\bibnamefont {van~der
  Zant}},\ }\href {\doibase https://doi.org/10.1016/S0370-1573(01)00022-9}
  {\bibfield  {journal} {\bibinfo  {journal} {Physics Reports}\ }\textbf
  {\bibinfo {volume} {355}},\ \bibinfo {pages} {235 } (\bibinfo {year}
  {2001})}\BibitemShut {NoStop}%
\bibitem [{\citenamefont {Machida}\ and\ \citenamefont
  {Koyama}(2013)}]{MACHIDA201344}%
  \BibitemOpen
  \bibfield  {author} {\bibinfo {author} {\bibfnamefont {K.}~\bibnamefont
  {Machida}, \bibfnamefont {M.and~Kobayashi}}\ and\ \bibinfo {author}
  {\bibfnamefont {T.}~\bibnamefont {Koyama}},\ }\href {\doibase
  10.1016/j.physc.2013.02.004} {\bibfield  {journal} {\bibinfo  {journal}
  {Phys. C}\ }\textbf {\bibinfo {volume} {491}},\ \bibinfo {pages} {44–46}
  (\bibinfo {year} {2013})}\BibitemShut {NoStop}%
\bibitem [{\citenamefont {Balents}\ \emph {et~al.}(2005)\citenamefont
  {Balents}, \citenamefont {Bartosch}, \citenamefont {Burkov}, \citenamefont
  {Sachdev},\ and\ \citenamefont {Sengupta}}]{PhysRevB.71.144508}%
  \BibitemOpen
  \bibfield  {author} {\bibinfo {author} {\bibfnamefont {L.}~\bibnamefont
  {Balents}}, \bibinfo {author} {\bibfnamefont {L.}~\bibnamefont {Bartosch}},
  \bibinfo {author} {\bibfnamefont {A.}~\bibnamefont {Burkov}}, \bibinfo
  {author} {\bibfnamefont {S.}~\bibnamefont {Sachdev}}, \ and\ \bibinfo
  {author} {\bibfnamefont {K.}~\bibnamefont {Sengupta}},\ }\href {\doibase
  10.1103/PhysRevB.71.144508} {\bibfield  {journal} {\bibinfo  {journal} {Phys.
  Rev. B}\ }\textbf {\bibinfo {volume} {71}},\ \bibinfo {pages} {144508}
  (\bibinfo {year} {2005})}\BibitemShut {NoStop}%
\bibitem [{\citenamefont {Hoffman}\ \emph {et~al.}(2002)\citenamefont
  {Hoffman}, \citenamefont {Hudson}, \citenamefont {Lang}, \citenamefont
  {Madhavan}, \citenamefont {Eisaki}, \citenamefont {Uchida},\ and\
  \citenamefont {Davis}}]{Hoffman466}%
  \BibitemOpen
  \bibfield  {author} {\bibinfo {author} {\bibfnamefont {J.~E.}\ \bibnamefont
  {Hoffman}}, \bibinfo {author} {\bibfnamefont {E.~W.}\ \bibnamefont {Hudson}},
  \bibinfo {author} {\bibfnamefont {K.~M.}\ \bibnamefont {Lang}}, \bibinfo
  {author} {\bibfnamefont {V.}~\bibnamefont {Madhavan}}, \bibinfo {author}
  {\bibfnamefont {H.}~\bibnamefont {Eisaki}}, \bibinfo {author} {\bibfnamefont
  {S.}~\bibnamefont {Uchida}}, \ and\ \bibinfo {author} {\bibfnamefont {J.~C.}\
  \bibnamefont {Davis}},\ }\href {\doibase 10.1126/science.1066974} {\bibfield
  {journal} {\bibinfo  {journal} {Science}\ }\textbf {\bibinfo {volume}
  {295}},\ \bibinfo {pages} {466} (\bibinfo {year} {2002})}\BibitemShut
  {NoStop}%
\bibitem [{\citenamefont {Preiss}()}]{Preiss-phdthesis}%
  \BibitemOpen
  \bibfield  {author} {\bibinfo {author} {\bibfnamefont {P.~M.}\ \bibnamefont
  {Preiss}},\ }\emph {\bibinfo {title} {Atomic Bose-Hubbard Systems with
  Single-Particle Control}},\ \href@noop {} {Ph.D. thesis},\ \bibinfo  {school}
  {Harvard University}\BibitemShut {NoStop}%
\bibitem [{\citenamefont {Ueda}(2010)}]{Ueda_book}%
  \BibitemOpen
  \bibfield  {author} {\bibinfo {author} {\bibfnamefont {M.}~\bibnamefont
  {Ueda}},\ }\href@noop {} {\emph {\bibinfo {title} {Fundamentals and New
  Frontiers of Bose-Einstein Condensation}}},\ \bibinfo {edition} {1st}\ ed.\
  (\bibinfo  {publisher} {World Scientific},\ \bibinfo {year}
  {2010})\BibitemShut {NoStop}%
\bibitem [{\citenamefont {Zhang}\ and\ \citenamefont {Dong}(2010)}]{ED}%
  \BibitemOpen
  \bibfield  {author} {\bibinfo {author} {\bibfnamefont {J.~M.}\ \bibnamefont
  {Zhang}}\ and\ \bibinfo {author} {\bibfnamefont {R.~X.}\ \bibnamefont
  {Dong}},\ }\href {http://stacks.iop.org/0143-0807/31/i=3/a=016} {\bibfield
  {journal} {\bibinfo  {journal} {Eur. J. Phys.}\ }\textbf {\bibinfo {volume}
  {31}},\ \bibinfo {pages} {591} (\bibinfo {year} {2010})}\BibitemShut
  {NoStop}%
\bibitem [{\citenamefont {Dirac}(1931)}]{Dirac60}%
  \BibitemOpen
  \bibfield  {author} {\bibinfo {author} {\bibfnamefont {P.~A.~M.}\
  \bibnamefont {Dirac}},\ }\href {\doibase 10.1098/rspa.1931.0130} {\bibfield
  {journal} {\bibinfo  {journal} {Proc. R. Soc. London, Ser. A}\ }\textbf
  {\bibinfo {volume} {133}},\ \bibinfo {pages} {60} (\bibinfo {year}
  {1931})}\BibitemShut {NoStop}%
\bibitem [{\citenamefont {Nielsen}\ and\ \citenamefont
  {Chuang}(2000)}]{Nielsen}%
  \BibitemOpen
  \bibfield  {author} {\bibinfo {author} {\bibfnamefont {M.~A.}\ \bibnamefont
  {Nielsen}}\ and\ \bibinfo {author} {\bibfnamefont {I.~L.}\ \bibnamefont
  {Chuang}},\ }\href@noop {} {\emph {\bibinfo {title} {Quantum Computation and
  Quantum information}}},\ \bibinfo {edition} {1st}\ ed.\ (\bibinfo
  {publisher} {Cambridge University Press},\ \bibinfo {year}
  {2000})\BibitemShut {NoStop}%
\bibitem [{\citenamefont {Wootters}(1998)}]{Wootters_etg}%
  \BibitemOpen
  \bibfield  {author} {\bibinfo {author} {\bibfnamefont {W.~K.}\ \bibnamefont
  {Wootters}},\ }\href {\doibase 10.1103/PhysRevLett.80.2245} {\bibfield
  {journal} {\bibinfo  {journal} {Phys. Rev. Lett.}\ }\textbf {\bibinfo
  {volume} {80}},\ \bibinfo {pages} {2245} (\bibinfo {year}
  {1998})}\BibitemShut {NoStop}%
\bibitem [{\citenamefont {Gerbier}\ \emph
  {et~al.}(2005{\natexlab{a}})\citenamefont {Gerbier}, \citenamefont {Widera},
  \citenamefont {F\"olling}, \citenamefont {Mandel}, \citenamefont {Gericke},\
  and\ \citenamefont {Bloch}}]{PhysRevLett.95.050404}%
  \BibitemOpen
  \bibfield  {author} {\bibinfo {author} {\bibfnamefont {F.}~\bibnamefont
  {Gerbier}}, \bibinfo {author} {\bibfnamefont {A.}~\bibnamefont {Widera}},
  \bibinfo {author} {\bibfnamefont {S.}~\bibnamefont {F\"olling}}, \bibinfo
  {author} {\bibfnamefont {O.}~\bibnamefont {Mandel}}, \bibinfo {author}
  {\bibfnamefont {T.}~\bibnamefont {Gericke}}, \ and\ \bibinfo {author}
  {\bibfnamefont {I.}~\bibnamefont {Bloch}},\ }\href {\doibase
  10.1103/PhysRevLett.95.050404} {\bibfield  {journal} {\bibinfo  {journal}
  {Phys. Rev. Lett.}\ }\textbf {\bibinfo {volume} {95}},\ \bibinfo {pages}
  {050404} (\bibinfo {year} {2005}{\natexlab{a}})}\BibitemShut {NoStop}%
\bibitem [{\citenamefont {Gerbier}\ \emph
  {et~al.}(2005{\natexlab{b}})\citenamefont {Gerbier}, \citenamefont {Widera},
  \citenamefont {F\"olling}, \citenamefont {Mandel}, \citenamefont {Gericke},\
  and\ \citenamefont {Bloch}}]{PhysRevA.72.053606}%
  \BibitemOpen
  \bibfield  {author} {\bibinfo {author} {\bibfnamefont {F.}~\bibnamefont
  {Gerbier}}, \bibinfo {author} {\bibfnamefont {A.}~\bibnamefont {Widera}},
  \bibinfo {author} {\bibfnamefont {S.}~\bibnamefont {F\"olling}}, \bibinfo
  {author} {\bibfnamefont {O.}~\bibnamefont {Mandel}}, \bibinfo {author}
  {\bibfnamefont {T.}~\bibnamefont {Gericke}}, \ and\ \bibinfo {author}
  {\bibfnamefont {I.}~\bibnamefont {Bloch}},\ }\href {\doibase
  10.1103/PhysRevA.72.053606} {\bibfield  {journal} {\bibinfo  {journal} {Phys.
  Rev. A}\ }\textbf {\bibinfo {volume} {72}},\ \bibinfo {pages} {053606}
  (\bibinfo {year} {2005}{\natexlab{b}})}\BibitemShut {NoStop}%
\bibitem [{\citenamefont {Hoffmann}\ and\ \citenamefont
  {Pelster}(2009)}]{PhysRevA.79.053623}%
  \BibitemOpen
  \bibfield  {author} {\bibinfo {author} {\bibfnamefont {A.}~\bibnamefont
  {Hoffmann}}\ and\ \bibinfo {author} {\bibfnamefont {A.}~\bibnamefont
  {Pelster}},\ }\href {\doibase 10.1103/PhysRevA.79.053623} {\bibfield
  {journal} {\bibinfo  {journal} {Phys. Rev. A}\ }\textbf {\bibinfo {volume}
  {79}},\ \bibinfo {pages} {053623} (\bibinfo {year} {2009})}\BibitemShut
  {NoStop}%
\bibitem [{\citenamefont {Zaleski}\ and\ \citenamefont
  {Kope\ifmmode~\acute{c}\else \'{c}\fi{}}(2011)}]{PhysRevA.84.053613}%
  \BibitemOpen
  \bibfield  {author} {\bibinfo {author} {\bibfnamefont {T.~A.}\ \bibnamefont
  {Zaleski}}\ and\ \bibinfo {author} {\bibfnamefont {T.~K.}\ \bibnamefont
  {Kope\ifmmode~\acute{c}\else \'{c}\fi{}}},\ }\href {\doibase
  10.1103/PhysRevA.84.053613} {\bibfield  {journal} {\bibinfo  {journal} {Phys.
  Rev. A}\ }\textbf {\bibinfo {volume} {84}},\ \bibinfo {pages} {053613}
  (\bibinfo {year} {2011})}\BibitemShut {NoStop}%
\bibitem [{\citenamefont {Polak}\ and\ \citenamefont
  {Zaleski}(2013)}]{PhysRevA.87.033614}%
  \BibitemOpen
  \bibfield  {author} {\bibinfo {author} {\bibfnamefont {T.~P.}\ \bibnamefont
  {Polak}}\ and\ \bibinfo {author} {\bibfnamefont {T.~A.}\ \bibnamefont
  {Zaleski}},\ }\href {\doibase 10.1103/PhysRevA.87.033614} {\bibfield
  {journal} {\bibinfo  {journal} {Phys. Rev. A}\ }\textbf {\bibinfo {volume}
  {87}},\ \bibinfo {pages} {033614} (\bibinfo {year} {2013})}\BibitemShut
  {NoStop}%
\bibitem [{\citenamefont {Islam}\ \emph {et~al.}(2015)\citenamefont {Islam},
  \citenamefont {Ma}, \citenamefont {Preiss}, \citenamefont {Eric~Tai},
  \citenamefont {Lukin}, \citenamefont {Rispoli},\ and\ \citenamefont
  {Greiner}}]{Greiner_etg}%
  \BibitemOpen
  \bibfield  {author} {\bibinfo {author} {\bibfnamefont {R.}~\bibnamefont
  {Islam}}, \bibinfo {author} {\bibfnamefont {R.}~\bibnamefont {Ma}}, \bibinfo
  {author} {\bibfnamefont {P.~M.}\ \bibnamefont {Preiss}}, \bibinfo {author}
  {\bibfnamefont {M.}~\bibnamefont {Eric~Tai}}, \bibinfo {author}
  {\bibfnamefont {A.}~\bibnamefont {Lukin}}, \bibinfo {author} {\bibfnamefont
  {M.}~\bibnamefont {Rispoli}}, \ and\ \bibinfo {author} {\bibfnamefont
  {M.}~\bibnamefont {Greiner}},\ }\href {http://dx.doi.org/10.1038/nature15750}
  {\bibfield  {journal} {\bibinfo  {journal} {Nature}\ }\textbf {\bibinfo
  {volume} {528}},\ \bibinfo {pages} {77} (\bibinfo {year} {2015})}\BibitemShut
  {NoStop}%
\bibitem [{Jpi()}]{Jpin_deg}%
  \BibitemOpen
  \href@noop {} {}\bibinfo {note} {Some cases have doubly degenerat ground
  states (e.g. $N_{\Phi}=1,3$), for which the entanglement is minimized with
  respect to all superpositions of the ground states. It turns out that the
  entanglement is insusceptible to different superpositions.}\BibitemShut
  {Stop}%
\bibitem [{\citenamefont {Wu}\ and\ \citenamefont {Lidar}(2002)}]{parafermion}%
  \BibitemOpen
  \bibfield  {author} {\bibinfo {author} {\bibfnamefont {L.-A.}\ \bibnamefont
  {Wu}}\ and\ \bibinfo {author} {\bibfnamefont {D.~A.}\ \bibnamefont {Lidar}},\
  }\href {\doibase 10.1063/1.1499208} {\bibfield  {journal} {\bibinfo
  {journal} {J. Math. Phys.}\ }\textbf {\bibinfo {volume} {43}},\ \bibinfo
  {pages} {4506} (\bibinfo {year} {2002})}\BibitemShut {NoStop}%
\bibitem [{\citenamefont {Verstraete}\ and\ \citenamefont
  {Cirac}(2004)}]{PhysRevA.70.060302}%
  \BibitemOpen
  \bibfield  {author} {\bibinfo {author} {\bibfnamefont {F.}~\bibnamefont
  {Verstraete}}\ and\ \bibinfo {author} {\bibfnamefont {J.~I.}\ \bibnamefont
  {Cirac}},\ }\href {\doibase 10.1103/PhysRevA.70.060302} {\bibfield  {journal}
  {\bibinfo  {journal} {Phys. Rev. A}\ }\textbf {\bibinfo {volume} {70}},\
  \bibinfo {pages} {060302} (\bibinfo {year} {2004})}\BibitemShut {NoStop}%
\bibitem [{\citenamefont {Kasamatsu}\ \emph {et~al.}(2013)\citenamefont
  {Kasamatsu}, \citenamefont {Ichinose},\ and\ \citenamefont
  {Matsui}}]{PhysRevLett.111.115303}%
  \BibitemOpen
  \bibfield  {author} {\bibinfo {author} {\bibfnamefont {K.}~\bibnamefont
  {Kasamatsu}}, \bibinfo {author} {\bibfnamefont {I.}~\bibnamefont {Ichinose}},
  \ and\ \bibinfo {author} {\bibfnamefont {T.}~\bibnamefont {Matsui}},\ }\href
  {\doibase 10.1103/PhysRevLett.111.115303} {\bibfield  {journal} {\bibinfo
  {journal} {Phys. Rev. Lett.}\ }\textbf {\bibinfo {volume} {111}},\ \bibinfo
  {pages} {115303} (\bibinfo {year} {2013})}\BibitemShut {NoStop}%
\bibitem [{\citenamefont {Baier}\ \emph {et~al.}(2016)\citenamefont {Baier},
  \citenamefont {Mark}, \citenamefont {Petter}, \citenamefont {Aikawa},
  \citenamefont {Chomaz}, \citenamefont {Cai}, \citenamefont {Baranov},
  \citenamefont {Zoller},\ and\ \citenamefont {Ferlaino}}]{Baier201}%
  \BibitemOpen
  \bibfield  {author} {\bibinfo {author} {\bibfnamefont {S.}~\bibnamefont
  {Baier}}, \bibinfo {author} {\bibfnamefont {M.~J.}\ \bibnamefont {Mark}},
  \bibinfo {author} {\bibfnamefont {D.}~\bibnamefont {Petter}}, \bibinfo
  {author} {\bibfnamefont {K.}~\bibnamefont {Aikawa}}, \bibinfo {author}
  {\bibfnamefont {L.}~\bibnamefont {Chomaz}}, \bibinfo {author} {\bibfnamefont
  {Z.}~\bibnamefont {Cai}}, \bibinfo {author} {\bibfnamefont {M.}~\bibnamefont
  {Baranov}}, \bibinfo {author} {\bibfnamefont {P.}~\bibnamefont {Zoller}}, \
  and\ \bibinfo {author} {\bibfnamefont {F.}~\bibnamefont {Ferlaino}},\ }\href
  {\doibase 10.1126/science.aac9812} {\bibfield  {journal} {\bibinfo  {journal}
  {Science}\ }\textbf {\bibinfo {volume} {352}},\ \bibinfo {pages} {201}
  (\bibinfo {year} {2016})}\BibitemShut {NoStop}%
\bibitem [{\citenamefont {Kuno}\ \emph {et~al.}(2016)\citenamefont {Kuno},
  \citenamefont {Sakane}, \citenamefont {Kasamatsu}, \citenamefont {Ichinose},\
  and\ \citenamefont {Matsui}}]{PhysRevA.94.063641}%
  \BibitemOpen
  \bibfield  {author} {\bibinfo {author} {\bibfnamefont {Y.}~\bibnamefont
  {Kuno}}, \bibinfo {author} {\bibfnamefont {S.}~\bibnamefont {Sakane}},
  \bibinfo {author} {\bibfnamefont {K.}~\bibnamefont {Kasamatsu}}, \bibinfo
  {author} {\bibfnamefont {I.}~\bibnamefont {Ichinose}}, \ and\ \bibinfo
  {author} {\bibfnamefont {T.}~\bibnamefont {Matsui}},\ }\href {\doibase
  10.1103/PhysRevA.94.063641} {\bibfield  {journal} {\bibinfo  {journal} {Phys.
  Rev. A}\ }\textbf {\bibinfo {volume} {94}},\ \bibinfo {pages} {063641}
  (\bibinfo {year} {2016})}\BibitemShut {NoStop}%
\bibitem [{\citenamefont {González-Cuadra}\ \emph {et~al.}(2017)\citenamefont
  {González-Cuadra}, \citenamefont {Zohar},\ and\ \citenamefont
  {Cirac}}]{Cirac_NJP}%
  \BibitemOpen
  \bibfield  {author} {\bibinfo {author} {\bibfnamefont {D.}~\bibnamefont
  {González-Cuadra}}, \bibinfo {author} {\bibfnamefont {E.}~\bibnamefont
  {Zohar}}, \ and\ \bibinfo {author} {\bibfnamefont {J.~I.}\ \bibnamefont
  {Cirac}},\ }\href {http://stacks.iop.org/1367-2630/19/i=6/a=063038}
  {\bibfield  {journal} {\bibinfo  {journal} {New J. Phys.}\ }\textbf {\bibinfo
  {volume} {19}},\ \bibinfo {pages} {063038} (\bibinfo {year}
  {2017})}\BibitemShut {NoStop}%
\bibitem [{\citenamefont {Cole}\ \emph {et~al.}(2012)\citenamefont {Cole},
  \citenamefont {Zhang}, \citenamefont {Paramekanti},\ and\ \citenamefont
  {Trivedi}}]{PhysRevLett.109.085302}%
  \BibitemOpen
  \bibfield  {author} {\bibinfo {author} {\bibfnamefont {W.~S.}\ \bibnamefont
  {Cole}}, \bibinfo {author} {\bibfnamefont {S.}~\bibnamefont {Zhang}},
  \bibinfo {author} {\bibfnamefont {A.}~\bibnamefont {Paramekanti}}, \ and\
  \bibinfo {author} {\bibfnamefont {N.}~\bibnamefont {Trivedi}},\ }\href
  {\doibase 10.1103/PhysRevLett.109.085302} {\bibfield  {journal} {\bibinfo
  {journal} {Phys. Rev. Lett.}\ }\textbf {\bibinfo {volume} {109}},\ \bibinfo
  {pages} {085302} (\bibinfo {year} {2012})}\BibitemShut {NoStop}%
\bibitem [{\citenamefont {Massignan}\ \emph {et~al.}(2010)\citenamefont
  {Massignan}, \citenamefont {Sanpera},\ and\ \citenamefont
  {Lewenstein}}]{PhysRevA.81.031607}%
  \BibitemOpen
  \bibfield  {author} {\bibinfo {author} {\bibfnamefont {P.}~\bibnamefont
  {Massignan}}, \bibinfo {author} {\bibfnamefont {A.}~\bibnamefont {Sanpera}},
  \ and\ \bibinfo {author} {\bibfnamefont {M.}~\bibnamefont {Lewenstein}},\
  }\href {\doibase 10.1103/PhysRevA.81.031607} {\bibfield  {journal} {\bibinfo
  {journal} {Phys. Rev. A}\ }\textbf {\bibinfo {volume} {81}},\ \bibinfo
  {pages} {031607} (\bibinfo {year} {2010})}\BibitemShut {NoStop}%
\bibitem [{\citenamefont {Huang}\ \emph {et~al.}(2016)\citenamefont {Huang},
  \citenamefont {Meng}, \citenamefont {Wang}, \citenamefont {Peng},
  \citenamefont {Zhang}, \citenamefont {Chen}, \citenamefont {Li},
  \citenamefont {Zhou},\ and\ \citenamefont {Zhang}}]{Huang_Fermi}%
  \BibitemOpen
  \bibfield  {author} {\bibinfo {author} {\bibfnamefont {L.}~\bibnamefont
  {Huang}}, \bibinfo {author} {\bibfnamefont {Z.}~\bibnamefont {Meng}},
  \bibinfo {author} {\bibfnamefont {P.}~\bibnamefont {Wang}}, \bibinfo {author}
  {\bibfnamefont {P.}~\bibnamefont {Peng}}, \bibinfo {author} {\bibfnamefont
  {S.-L.}\ \bibnamefont {Zhang}}, \bibinfo {author} {\bibfnamefont
  {L.}~\bibnamefont {Chen}}, \bibinfo {author} {\bibfnamefont {D.}~\bibnamefont
  {Li}}, \bibinfo {author} {\bibfnamefont {Q.}~\bibnamefont {Zhou}}, \ and\
  \bibinfo {author} {\bibfnamefont {J.}~\bibnamefont {Zhang}},\ }\href
  {http://dx.doi.org/10.1038/nphys3672} {\bibfield  {journal} {\bibinfo
  {journal} {Nature Phys.}\ }\textbf {\bibinfo {volume} {12}},\ \bibinfo
  {pages} {540} (\bibinfo {year} {2016})}\BibitemShut {NoStop}%
\bibitem [{\citenamefont {Fedorov}\ \emph {et~al.}(2017)\citenamefont
  {Fedorov}, \citenamefont {Yudson},\ and\ \citenamefont
  {Shlyapnikov}}]{PhysRevA.95.043615}%
  \BibitemOpen
  \bibfield  {author} {\bibinfo {author} {\bibfnamefont {A.~K.}\ \bibnamefont
  {Fedorov}}, \bibinfo {author} {\bibfnamefont {V.~I.}\ \bibnamefont {Yudson}},
  \ and\ \bibinfo {author} {\bibfnamefont {G.~V.}\ \bibnamefont
  {Shlyapnikov}},\ }\href {\doibase 10.1103/PhysRevA.95.043615} {\bibfield
  {journal} {\bibinfo  {journal} {Phys. Rev. A}\ }\textbf {\bibinfo {volume}
  {95}},\ \bibinfo {pages} {043615} (\bibinfo {year} {2017})}\BibitemShut
  {NoStop}%
\bibitem [{\citenamefont {Sachdev}(2011)}]{Sachdev_book}%
  \BibitemOpen
  \bibfield  {author} {\bibinfo {author} {\bibfnamefont {S.}~\bibnamefont
  {Sachdev}},\ }\href@noop {} {\emph {\bibinfo {title} {Quantum Phase
  Transitions}}},\ \bibinfo {edition} {2nd}\ ed.\ (\bibinfo  {publisher}
  {Cambridge University Press},\ \bibinfo {year} {2011})\BibitemShut {NoStop}%
\bibitem [{\citenamefont {\ifmmode~\dot{Z}\else \.{Z}\fi{}ukowski}\ \emph
  {et~al.}(1993)\citenamefont {\ifmmode~\dot{Z}\else \.{Z}\fi{}ukowski},
  \citenamefont {Zeilinger}, \citenamefont {Horne},\ and\ \citenamefont
  {Ekert}}]{PhysRevLett.71.4287}%
  \BibitemOpen
  \bibfield  {author} {\bibinfo {author} {\bibfnamefont {M.}~\bibnamefont
  {\ifmmode~\dot{Z}\else \.{Z}\fi{}ukowski}}, \bibinfo {author} {\bibfnamefont
  {A.}~\bibnamefont {Zeilinger}}, \bibinfo {author} {\bibfnamefont {M.~A.}\
  \bibnamefont {Horne}}, \ and\ \bibinfo {author} {\bibfnamefont {A.~K.}\
  \bibnamefont {Ekert}},\ }\href {\doibase 10.1103/PhysRevLett.71.4287}
  {\bibfield  {journal} {\bibinfo  {journal} {Phys. Rev. Lett.}\ }\textbf
  {\bibinfo {volume} {71}},\ \bibinfo {pages} {4287} (\bibinfo {year}
  {1993})}\BibitemShut {NoStop}%
\bibitem [{\citenamefont {Auerbach}(1994)}]{Auerbach_book}%
  \BibitemOpen
  \bibfield  {author} {\bibinfo {author} {\bibfnamefont {A.}~\bibnamefont
  {Auerbach}},\ }\href@noop {} {\emph {\bibinfo {title} {Interacting Electrons
  and Quantum Magnetism}}}\ (\bibinfo  {publisher} {Springer-Verlag},\ \bibinfo
  {address} {UK},\ \bibinfo {year} {1994})\BibitemShut {NoStop}%
\bibitem [{\citenamefont {Blais}\ \emph {et~al.}(2004)\citenamefont {Blais},
  \citenamefont {Huang}, \citenamefont {Wallraff}, \citenamefont {Girvin},\
  and\ \citenamefont {Schoelkopf}}]{PhysRevA.69.062320}%
  \BibitemOpen
  \bibfield  {author} {\bibinfo {author} {\bibfnamefont {A.}~\bibnamefont
  {Blais}}, \bibinfo {author} {\bibfnamefont {R.-S.}\ \bibnamefont {Huang}},
  \bibinfo {author} {\bibfnamefont {A.}~\bibnamefont {Wallraff}}, \bibinfo
  {author} {\bibfnamefont {S.~M.}\ \bibnamefont {Girvin}}, \ and\ \bibinfo
  {author} {\bibfnamefont {R.~J.}\ \bibnamefont {Schoelkopf}},\ }\href
  {\doibase 10.1103/PhysRevA.69.062320} {\bibfield  {journal} {\bibinfo
  {journal} {Phys. Rev. A}\ }\textbf {\bibinfo {volume} {69}},\ \bibinfo
  {pages} {062320} (\bibinfo {year} {2004})}\BibitemShut {NoStop}%
\bibitem [{\citenamefont {Nigg}\ and\ \citenamefont
  {Girvin}(2013)}]{PhysRevLett.110.243604}%
  \BibitemOpen
  \bibfield  {author} {\bibinfo {author} {\bibfnamefont {S.~E.}\ \bibnamefont
  {Nigg}}\ and\ \bibinfo {author} {\bibfnamefont {S.~M.}\ \bibnamefont
  {Girvin}},\ }\href {\doibase 10.1103/PhysRevLett.110.243604} {\bibfield
  {journal} {\bibinfo  {journal} {Phys. Rev. Lett.}\ }\textbf {\bibinfo
  {volume} {110}},\ \bibinfo {pages} {243604} (\bibinfo {year}
  {2013})}\BibitemShut {NoStop}%
\bibitem [{\citenamefont {Dai}\ \emph {et~al.}(2015)\citenamefont {Dai},
  \citenamefont {Kuo},\ and\ \citenamefont {Chung}}]{srep_Dai}%
  \BibitemOpen
  \bibfield  {author} {\bibinfo {author} {\bibfnamefont {L.}~\bibnamefont
  {Dai}}, \bibinfo {author} {\bibfnamefont {W.}~\bibnamefont {Kuo}}, \ and\
  \bibinfo {author} {\bibfnamefont {M.-C.}\ \bibnamefont {Chung}},\ }\href
  {\doibase 10.1038/srep11188} {\bibfield  {journal} {\bibinfo  {journal} {Sci.
  Rep.}\ }\textbf {\bibinfo {volume} {5}},\ \bibinfo {pages} {11188} (\bibinfo
  {year} {2015})}\BibitemShut {NoStop}%
\end{thebibliography}%
%\begin{thebibliography}{99}

% use separate files for references.

%\end{thebibliography}

\end{document}